\documentclass[twocolumn]{autart}    % Enable this line and disable the
\usepackage{graphicx}          % Include this line if your
\usepackage{amsmath}

\usepackage{amsfonts}
\usepackage{color}
\usepackage{epstopdf} % for postscript graphics files
\usepackage{tikz}
\usepackage{mathtools}
\newcommand{\smallsub}[1]{\! {\scriptscriptstyle \mathcal{#1}} }

\newtheorem{proposition}{Proposition}

\newtheorem{remark}{Remark}
\newtheorem{lemma}{Lemma}

\newtheorem{condition}{Condition}

\newcommand{\mY}{\mathcal{Y}}
\newcommand{\mD}{\mathcal{D}}
\newcommand{\mZ}{\mathcal{Z}}
\newcommand{\mQ}{\mathcal{Q}}

\newcommand{\mU}{\mathcal{U}}

\newcommand{\mL}{\mathcal{L}}
\newcommand{\mX}{\mathcal{X}}
\newcommand{\mO}{\mathcal{O}}

\newcommand{\smO}{\smallsub{O}}
\newcommand{\smU}{\smallsub{U}}
\newcommand{\smZ}{\smallsub{Z}}

\newcommand{\smC}{\smallsub{C}}
\newcommand{\smD}{\smallsub{D}}
\newcommand{\smY}{\smallsub{Y}}
\newcommand{\smQ}{\smallsub{Q}}
\newcommand{\smX}{\smallsub{X}}
\newcommand{\smL}{\smallsub{L}}

\newcommand{\beq}{\begin{equation}}
\newcommand{\eeq}{\end{equation}}
\newcommand{\beqs}{\begin{equation*}}
\newcommand{\eeqs}{\end{equation*}}
\newcommand{\beqr}{\begin{eqnarray}}
\newcommand{\eeqr}{\end{eqnarray}}
\newcommand{\beqrs}{\begin{eqnarray*}}
	\newcommand{\eeqrs}{\end{eqnarray*}}

\newcommand{\mW}{\mathcal{W}}

\newcommand{\smW}{\smallsub{W}}

\DeclareMathOperator*{\argmin}{arg\,min}
\DeclareMathOperator*{\argmax}{arg\,max}
\usepackage{algorithm}
\newcommand{\norm}[1]{\| #1 \|}

\usepackage{cite}

\begin{document}

\begin{frontmatter}

\title{Learning linear modules in a dynamic network with missing node observations\thanksref{footnoteinfo}} % Title, preferably not more

\thanks[footnoteinfo]{Draft paper submitted to Automatica, 23 August 2022. This project has received funding from the European Research Council (ERC), Advanced Research Grant SYSDYNET, under the European Union’s Horizon 2020 research and innovation programme (Grant Agreement No. 694504).}

\author[TUE]{Karthik R. Ramaswamy}\ead{k.r.ramaswamy@tue.nl},    % Add the
\author[TUE]{Giulio Bottegal}\ead{giulio.bottegal@gmail.com},               % e-mail address
\author[TUE]{Paul M.J. Van den Hof}\ead{p.m.j.vandenhof@tue.nl}  % (ead) as shown

\address[TUE]{Department of Electrical Engineering, Eindhoven University of Technology, Eindhoven, The Netherlands}
%\address[ASML]{ASML, Veldhoven, The Netherlands}
%% Please supply
%\address[Paestum]{Senate House, Rome}             % full addresses
%\address[Paestum]{The White House, Baiae}        % here.

                             % keyword list or with the
                                          % help of the Automatica
                                          % keyword wizard

\begin{abstract}
In order to identify a system (module) embedded in a dynamic network, one has to formulate a multiple-input estimation problem that necessitates certain nodes to be measured and included as predictor inputs. However, some of these nodes may not be measurable in many practical cases due to sensor selection and placement issues. This may result in biased estimates of the target module. Furthermore, the identification problem associated with the multiple-input structure may require determining a large number of parameters that are not of particular interest to the experimenter, with increased computational complexity in large-sized networks. In this paper, we tackle these problems by using a data augmentation strategy that allows us to reconstruct the missing node measurements and increase the accuracy of the estimated target module. To this end, we develop a system identification method using regularized kernel-based methods coupled with approximate inference methods. Keeping a parametric model for the module of interest, we model the other modules as Gaussian Processes (GP) with a kernel given by the so-called stable spline kernel. An Empirical Bayes (EB) approach is used to estimate the parameters of the target module. The related optimization problem is solved using an Expectation-Maximization (EM) method, where we employ a Markov-chain Monte Carlo (MCMC) technique to reconstruct the unknown missing node information and the network dynamics. Numerical simulations on dynamic network examples illustrate the potentials of the developed method.
\end{abstract}

\end{frontmatter}

%%%%%%%%%%%%%%%%%%%%%%%%%%%%%%%%%%%%%%%%%%%%%%%%%%%%%%%%%%%%%%%%%%%%%%%%%%%%%%%%
\section{Introduction}
Modern systems are becoming increasingly complex and largely interconnected, which is evident from the growing interest in the field of data-driven modeling in dynamic networks. Dynamic networks are typically represented as a set of measurable signals (node signals) interconnected through linear dynamic systems (modules), that are possibly driven by external excitation signals. Data-driven model learning in dynamic networks can be classified into two categories - full network identification and local module identification. The former deals with the identification of full network dynamics \cite{Haber&Verhaegen:TAC:14,Torres14,Weerts&etal_Autom:18_reducedrank,Zorzi&Chiuso:17,Fonken&etal:Arxiv_AUT:21},
% including identifiability aspects \cite{Goncalves&Warnick:08, Weerts&etal_Autom:18_identifiability,Bazanella&etal_CDC:17,vanWaarde&Tesi&Camlibel_NECSYS:18,Cheng&etal_CDC:19},
and the latter deals with learning a single module embedded in a dynamic network given the topology of the network \cite{Chiuso&Pillonetto_Autom:12,VandenHof&etal_Autom:13,materassi2015identification,Dankers&etal_TAC:16,Linder&Enqvist_ijc:17,Ramaswamy&etal_CDC:18, Everitt&Bottegal&Hjalmarsson_Autom:18,Gevers&etal:sysid18,Ramaswamy&etal_CDC:19,Ramaswamy&etal_TAC:21,Ramaswamy&etal_Autom:2021,Yu&Verhaegen_TAC:2018,Ramaswamy&etal_Autom:22}. A brief survey on the local module identification has been presented in \cite{vandenhof&ramaswamy_PMLR:21}.

In this paper we focus on learning local modules in a dynamic network. In \cite{VandenHof&etal_Autom:13, Dankers&etal_TAC:16}, the classical closed loop identification methods \cite{Ljung:99} like \emph{direct method}, \emph{two-stage method} and \emph{joint input-output method} have been generalized to a dynamic network setting. A \emph{local direct method} that can deal with correlated noise has been introduced in \cite{Ramaswamy&etal_TAC:21}. An \emph{indirect method} to identify local modules has been presented in \cite{Gevers&etal:sysid18}. Apart from the aforementioned prediction error methods, the direct method and the two-stage method have been extended to a Bayesian setting using regularized kernel-based methods in \cite{Ramaswamy&etal_CDC:18,Ramaswamy&etal_Autom:2021} and \cite{Everitt&Bottegal&Hjalmarsson_Autom:18} respectively. A \emph{generalized method} that combines direct and indirect approaches to relax the sensing and excitation schemes has been introduced in \cite{Ramaswamy&etal_CDC:19}.
The above mentioned direct methods require solving a MISO or MIMO identification problem, resulting in the problem of estimating a large number of parameters that are of no prime interest to the experimenter and the associated optimization problem may become computationally infeasible, being nonlinear and large-dimensional. Also, it may be required to perform a preliminary model order selection step for each of the MISO modules, e.g. using complexity criteria such as AIC, BIC, or cross validation \cite{Ljung:99}. For a relatively large network, one has to test a huge number of combination of candidate model orders, which can become computationally infeasible already in medium-sized networks. For example, for 5 modules in the MISO setup with FIR model structure and orders from 1 to 5, we need to test $5^5$ possible combinations. These problems have been tackled in \cite{Ramaswamy&etal_CDC:18, Ramaswamy&etal_Autom:2021,Rajagopal&etal_CDC:20} using regularized kernel-based methods by extending the direct method framework to a Bayesian setting.

When using the direct method for single module identification, the fundamental \emph{parallel path/loop} condition \cite{Dankers&etal_TAC:16} needs to hold in order to achieve consistent estimates of the target module. This condition states that all parallel paths from the input to the output of the target module and all the loops around the output of the target module must pass through nodes that are measured and included as predictor inputs in the estimation problem. Therefore, it becomes quintessential to measure certain nodes to satisfy the \emph{parallel path/loop} condition; however, measurement of such nodes might not always be possible. Therefore, to mitigate this issue and achieve reduced bias estimates, it becomes essential to develop identification methods to cope with networks with missing node observations.

In this paper, we introduce a novel identification method that handles the situation of non-measured inputs (i.e., missing node observations) that are required to obtain unbiased target module estimates. To handle the situation of missing node observations, we use a \emph{data augmentation} strategy \cite{Tanner&Wong_ASA:1987, Vandyk&Meng_CGS:2001}, where the missing node observations are also estimated along with the parameters of the modules. For reconstructing the missing node information, we use the available information from nodes that lie both upstream and downstream compared to the missing node. To avoid model order selection and reduce the number of nuisance parameters to be estimated, we build on \cite{Ramaswamy&etal_CDC:18,Ramaswamy&etal_Autom:2021} and employ non-parametric regularized kernel-based methods.
We keep a parametric model only for the module of interest in order to have
an accurate description of its dynamics. The rest of the modules are modeled as zero-mean Gaussian processes, with covariance matrix (kernel) given by a \emph{first-order stable spline kernel} \cite{Pillonettoetal_Autom:14,Chenetal_Autom:12}, which encodes stability and smoothness of the processes. By this way of modeling, we reduce the number of estimated parameters and obtain a significant reduction in the variance of the estimates \cite{Pillonettoetal_Autom:14}.

% based on non-parametric regularized kernel-based methods \cite{Pillonettoetal_Autom:14} that improves the performance of the direct method:
%\begin{enumerate}
%	\item by handling the situation of non-measured inputs (i.e. missing node observations) that are required to obtain unbiased target module estimates;
%	\item avoids model order selection issues;
%	\item reduces the number of nuisance parameters to be estimated.
%\end{enumerate}
%To accomplish this, we develop the following approach. To handle the situation of missing node observations, we use a \emph{data augmentation} strategy \cite{Tanner&Wong_ASA:1987, Vandyk&Meng_CGS:2001}, where the missing node observations are also estimated along with the parameters of the modules. For reconstructing the missing node information, we use the available information from nodes that lie both upstream and downstream compared to the missing node. We keep a parametric model only for the module of interest in order to have
%an accurate description of its dynamics. The rest of the modules are modeled as zero-mean Gaussian processes, with covariance matrix (kernel) given by a \emph{first-order stable spline kernel} \cite{Pillonettoetal_Autom:14,Chenetal_Autom:12}, which encodes stability and smoothness of the processes. By this way of modeling, we reduce the number of parameters to be estimated and also obtain significant reduction in the variance of the estimates \cite{Pillonettoetal_Autom:14}.

Using the above approach, we obtain a Gaussian probabilistic description that depends on a vector of parameters that also contains the parameters of interest. We use an Empirical Bayes approach to estimate such a vector; this amounts to maximizing the marginal likelihood of the observed data, obtained by integrating out the dependence on the missing node data and the impulse response of the modules. The solution to the maximization problem is obtained using an iterative scheme based on the Expectation-Maximization (EM) \cite{Dempsteretal_JRSS:1977} algorithm. The E-step characterizing this scheme involves computing the expected value of a joint log-likelihood. Since in this problem the associated integral does not admit an analytical solution, we employ a Monte Carlo approximation method where samples of the target probability distributions are generated using an instance of the \emph{Gibbs sampler} \cite{Gemen&Gemen_PA&MI:1984}. As for the M-step of the EM procedure, we show that it can be split into several small optimization problems that are simple to solve, making the whole optimization routine computationally cheap. Numerical simulation on a dynamic network with missing node observations shows the developed method's potentials compared to the available classical methods.

This paper is organized as follows. In Section 2, the setup of the dynamic network and the problem statement is defined. Section 3 briefs about the standard direct method. Next, we provide the MIMO model, strategy to reduce the parameters of nuisance modules and solve the missing node observation problem, and solution to the marginal likelihood problem using the MCEM method in Section 4, 5, and 6. Next, numerical simulations and results are provided in Section 7, followed by conclusions. Finally, the technical proofs are provided in the appendix.

\section{Problem statement}\label{sec:prob stat}
Following the setup of \cite{VandenHof&etal_Autom:13}, a dynamic network is built up out of $L$ scalar \emph{internal variables} or \emph{nodes} $w_j$, $j = 1, \ldots, L$, and $K$ \emph{external variables} $r_k$, $k=1,\ldots K$.
Each internal variable is described as:
\begin{align}
w_j(t) = \sum_{\stackrel{l=1}{l\neq j}}^L
G_{jl}^0(q)w_l(t) + u_j(t) + v_j(t)
\label{eq:netw_def}
\end{align}
where $q^{-1}$ is the delay operator, i.e. $q^{-1}w_j(t) = w_j(t-1)$,
\begin{itemize}
	\item $G_{jl}^0$ are strictly proper rational transfer functions, referred to as {\it modules};
	\item There are no self-loops in the network, i.e. nodes are
	not directly connected to themselves ($G_{jj}^0 = 0$);
% 	\item $u_j(t)$ is generated by the \emph{external variables} $r_k(t)$ that can directly be manipulated by the user and is given by
% 	$u_j(t) = \sum_{k=1}^{K}R_{jk}r_k(t)$ where $R_{jk}$ are stable, proper rational transfer functions;
%	and that may or may not be present; if $r_j$ is not present it is replaced by $r_j=0$.
    \item $v_j(t)$ is the \emph{process noise}, where the vector process $v=[v_1 \cdots v_L]^T$ is modelled as a stationary stochastic process with rational spectral density $\Phi_v(\omega)$, such that there exists a Gaussian white noise process $e:= [e_1 \cdots e_L]^T$, with covariance matrix $\Lambda>0$ such that
	$ v(t) = H^0(q)e(t)$,
	where $H(q)$ is monic, square, stable and minimum-phase. The noises are uncorrelated i.e refers to the situation that $\Phi_v(\omega)$ and $H^0$ are diagonal.
	\item $u_j$ is an input signal, $u_j(t)=\sum_{k=1}^K R_{jk}^0(q)r_k(t)$ with $r_k$ being the \emph{known external signal} that can directly be manipulated by the user and $R^0_{jk}(q)$ is a known stable proper rational transfer function. In some nodes, it may be absent.
% 	\item $r_j(t)$ is a measured \emph{external excitation signal} entering node $w_j(t)$. In some nodes, it may be absent.
\end{itemize}

When combining the $L$ node signals we arrive at the full network expression
\begin{align*}
\begin{bmatrix}  \! w_1 \!  \\[1pt] \! w_2 \!  \\[1pt]  \! \vdots \! \\[1pt] \! w_L \!  \end{bmatrix} \! = \!
\begin{bmatrix}
0 &\! G_{12}^0 \!& \! \cdots \! &\!\! G_{1L}^0 \!\\
\! G_{21}^0 \!& 0 & \! \ddots \! &\!\!  \vdots \!\\
\vdots &\! \ddots \!& \! \ddots \! &\!\! G_{L-1 \ L}^0 \!\\
\! G_{L1}^0 \!&\! \cdots \!& \!\! G_{L \ L-1}^0 \!\! &\!\! 0
\end{bmatrix} \!\!
\begin{bmatrix} \! w_1 \!\\[1pt]  \! w_2 \!\\[1pt] \! \vdots \!\\[1pt] \! w_L \! \end{bmatrix} \!
+ \!
\begin{bmatrix} \! u_1 \!\\[1pt] \! u_2 \!\\[1pt] \! \vdots \!\\[1pt]  \! u_{L} \!\end{bmatrix}
\!+\!
H^0  \!\! \begin{bmatrix}\! e_1 \!\\[1pt] \! e_2 \!\\[1pt] \! \vdots \!\\[1pt] \! e_L\!\end{bmatrix} \!\!\!
\end{align*}
which results in the matrix equation:
\begin{align} \label{eq.dgsMatrix}
w = G^0 w + u + H^0 e.
\end{align}
It is assumed that the dynamic network is stable, i.e. $(I-G^0)^{-1}$ is stable, and well posed (see \cite{Dankers_diss:14} for details). The representation (\ref{eq.dgsMatrix}) is an extension of the dynamic structure function representation \cite{Goncalves&Warnick:08}.
The identification problem to be considered is the problem of identifying one particular module $G_{ji}^0(q)$ on the basis of a pre-specified subset of nodes $w$, and possibly $u$, assuming that $N$ samples of these variables have been observed and the network topology is known. To this end, we choose a parameterization of $G_{ji}^0(q)$, denoted as $G_{ji}(q,\theta)$, that describes the dynamics of the module of interest for a certain parameter vector $\theta = \theta_0 \in \mathbb{R}^{n_\theta}$.

\section{The direct method and predictor input selection}\label{sec:Direct_Id}
Let us define $\mathcal{N}_j$ as the set of node indices $k$ such that $G_{jk}^0 \neq 0$, i.e. the node signals in $\mathcal{N}_j$  are the \emph{$w$-in-neighbors of the node signal $w_j$}.
Following the definition of a dynamic network in the previous section, each scalar internal variable can be described as:
\begin{equation}\label{eq:singleblock}
 w_j(t) = \sum_{k \in \mathcal{N}_j} G_{jk}^0(q)w_k(t) + u_j(t) + v_j(t)
\end{equation}
%where $\mathcal{N}_j$ is the set of node indices $k$ such that $G_{jk} \neq 0$.
The above equation represents a MISO structure and is the starting point of the methodology presented in this paper, which is based on extending the MISO direct method \cite{VandenHof&etal_Autom:13}. In the standard MISO direct method for dynamic networks \cite{VandenHof&etal_Autom:13}, we consider the one-step-ahead predictor \cite{Ljung:99} of $w_j(t)$:
\begin{equation*}
	\begin{split}
	\hat w_j(t|t\!\!-\!\!1;\theta) \!\!=\! &\big(1\!\!-\!\!H_j^{-1}(q,\!\theta)\big)w_j(t) \!\!+\!\! H_j^{-1}(q,\!\theta) G_{ji}(q,\theta)w_i(t)\\
	&\!+\! H_j^{-1}(q,\!\theta)\big(\!\!\sum_{k \in \mathcal{N}_j\backslash\{i\}} \!\!G_{jk}(q,\theta)w_k(t) + u_j(t) \big)
	\end{split}
\end{equation*}
which is a function of the parameter vector $\theta$.
% Here $\mathcal{D}_j$ denote the set of indices of the internal variables that are chosen as predictor inputs. In order to identify an unbiased target module estimate using the direct method \cite{Dankers&etal_TAC:16}, the predictor inputs need to be chosen such that it satisfies the below condition.
% \begin{condition}[module invariance condition\cite{Dankers&etal_TAC:16}]
% \label{condx1}
% Let $G_{ji}$ be the target network module to be identified.
% In the original network (\ref{eq.dgsMatrix}):
% \begin{itemize}
% \item $i \in \mD_j$ and $j \notin \mD_j$;
% \item Every path from $w_i$ to $w_j$, excluding the path through $G_{ji}$, passes through a node $w_k, k \in \mD_j$, and
% \item Every loop through $w_j$ passes through a node in $w_k, k \in \mD_j$. \hfill $\Box$
% \end{itemize}
% \end{condition}

Not only the target module, but also the modules $G_{jk}^0(q)$, $k \in \mathcal{N}_j\backslash \{i\}$, and the noise model $H_j^0(q)$, are suitably parameterized with additional parameters. The parameter vector of interest $\theta$ is identified by minimizing the sum of the squared prediction error $\varepsilon_j(t) = w_j(t) - \hat w_j(t|t-1;\theta)$. We note that in this formulation, the prediction error depends also on the additional parameters entering the remaining modules and the noise model, which need to be identified to guarantee consistent estimates of $\theta$. Therefore, the total number of parameters may grow large if the cardinality of $\mathcal{N}_j$ is large, with a detrimental effect on the variance of the estimate of $\theta$ in the case where $N$ is not very large.

According to \cite{Dankers&etal_TAC:16}, it is sufficient to have a set of node signals $\mD_j$ to be measured and used as predictor inputs in the direct method, that satisfies an  additional parallel path/loop condition and a confounding variable condition.
\begin{condition}[parallel path and loop condition\cite{Dankers&etal_TAC:16}]
	\label{condx1}
	Let $G_{ji}$ be the target network module to be identified.
	In the network (\ref{eq.dgsMatrix}):
	\begin{itemize}
		\item Every path from $w_i$ to $w_j$, excluding the path through $G_{ji}$, passes through a node $w_k, k \in \mD_j$, and
		\item Every loop through $w_j$ passes through a node in $w_k, k \in \mD_j$. \hfill $\Box$
	\end{itemize}
\end{condition}
However, if $\mathcal{Z}_j \subseteq \mathcal{D}_j$ represent the node signals that are not measured or are inaccessible (i.e. \emph{missing node observations}) but required to satisfy the above condition, then identification through the direct method using the available signals leads to biased target module estimates \cite{Dankers&etal_TAC:16}. From this, we note that the direct method requires the measurement of the node signals $w_k, k \in \mD_j$.
% , but required to satisfy condition \eqref{condx1} to get unbiased estimates.
\begin{figure}
	\centering
	\hspace*{-0.5cm}
	\includegraphics[scale=0.5]{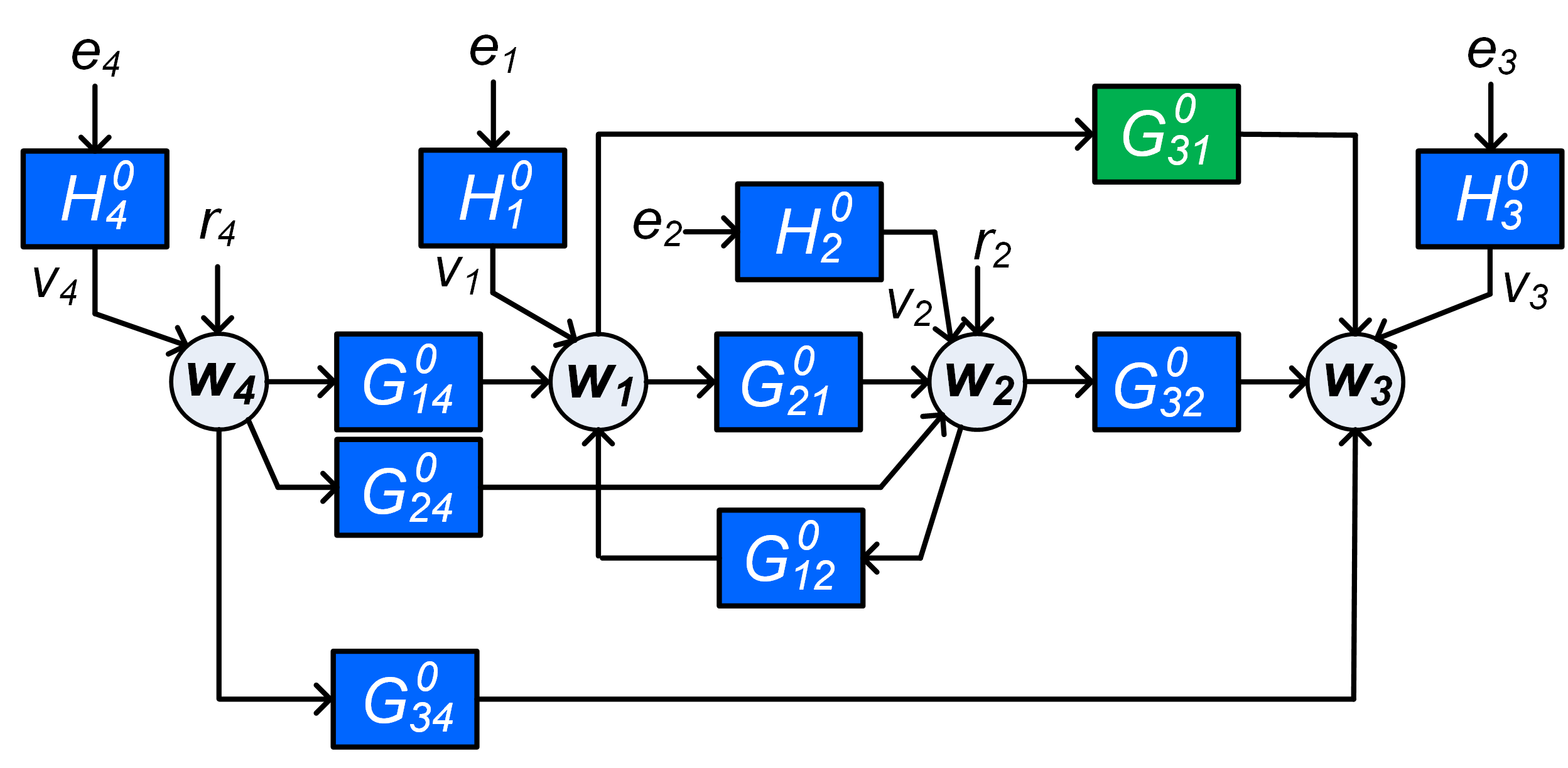}
	\caption{Network example with 4 internal nodes, 2 external excitation signals and a noise sources at each node. Target module is $G_{31}^0$.}
	\label{fig:dynnet_Ex_wnoise1_10}
\end{figure}
For example, consider the network in Figure \ref{fig:dynnet_Ex_wnoise1_10} with diagonal noise spectrum and $u_2 = r_2, u_4 = r_4$. In order to identify $G_{31}^0$ using the direct method, performing a MISO identification with $w_3$ as output and $w_{\smD_j} = \{w_1,w_4\}$ as inputs when $w_2$ is not measured (i.e. missing node observation), leads to estimation of modules in an immersed network (a network with $w_2$ removed) as in Figure \ref{fig:dynnet_Ex_wnoise2}. As we can see, we now estimate $G_{31}^0+G_{32}^0G_{21}^0$ (from $w_1$ to $w_3$) and not the desired target module $G_{31}^0$, which leads to a biased estimate. Confounding variables like $e_2$ when $w_2$ is non-measured also create bias in the estimate of the target module \cite{Ramaswamy&etal_TAC:21}.
\begin{figure}
	\centering
	\hspace*{-0.5cm}
	\includegraphics[scale=0.5]{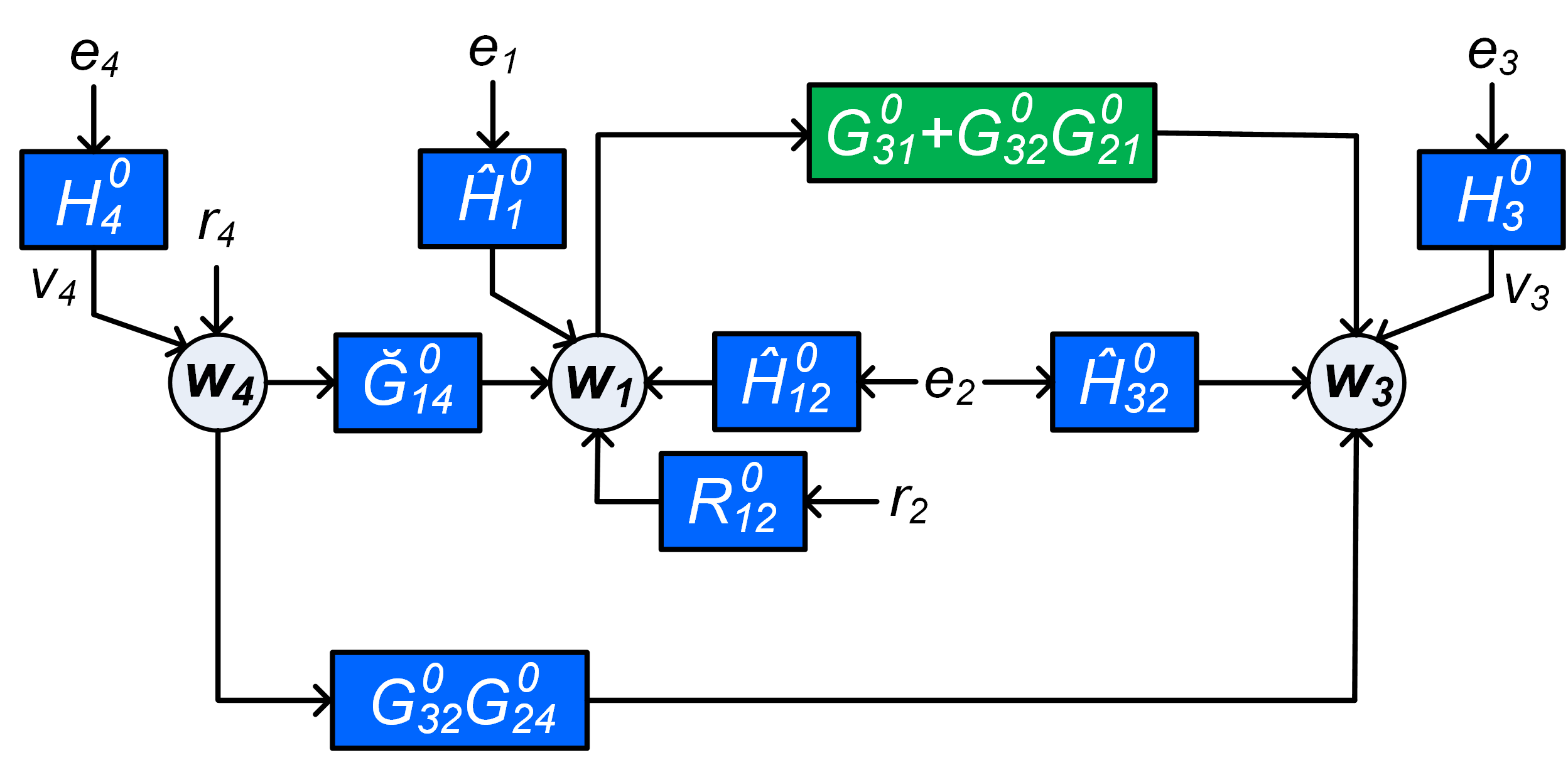}
	\caption{Immersed network of network in Figure \ref{fig:dynnet_Ex_wnoise1_10} where the non-measured node $w_2$ is removed.}
	\label{fig:dynnet_Ex_wnoise2}
\end{figure}
In the sequel of this paper, we will explain how we deal with the problem of missing node observations by re-constructing these measurements.

\section{Concepts and Notations}
We define the following concepts and notations.
\begin{defn}[confounding variable] \label{def1}
	Consider a dynamic network defined in \eqref{eq.dgsMatrix}
	with
	$e$ a white noise process such that $cov(e)$ being diagonal, and consider the graph related to this network, with node signals $w$  and $e$. Let $w_{\smX}$ and $w_{\smX'}$ be
	two subsets of node signals in $w$.
	Then a noise component $e_{\ell}$ in $e$ is a {\it confounding variable between $w_{\smX} \leftrightarrow w_{\smX'}$ given $w_{\smC}$}, if in the graph there exist simultaneous paths\footnote{A simultaneous path from $e_1$ to node signal $w_1$ and $w_2$ implies that there exist a path from $e_1$ to $w_1$ as well as from $e_1$ to $w_2$.} from $e_{\ell}$ to node signals $w_{k}, k \in \mX$ and $w_{n}, n \in \mX'$, while these paths do not run through a node in $w_{\smC}$. \hfill \qed
\end{defn}

\begin{figure}
	\centering
	\hspace*{-0.5cm}
	\includegraphics[scale=0.45]{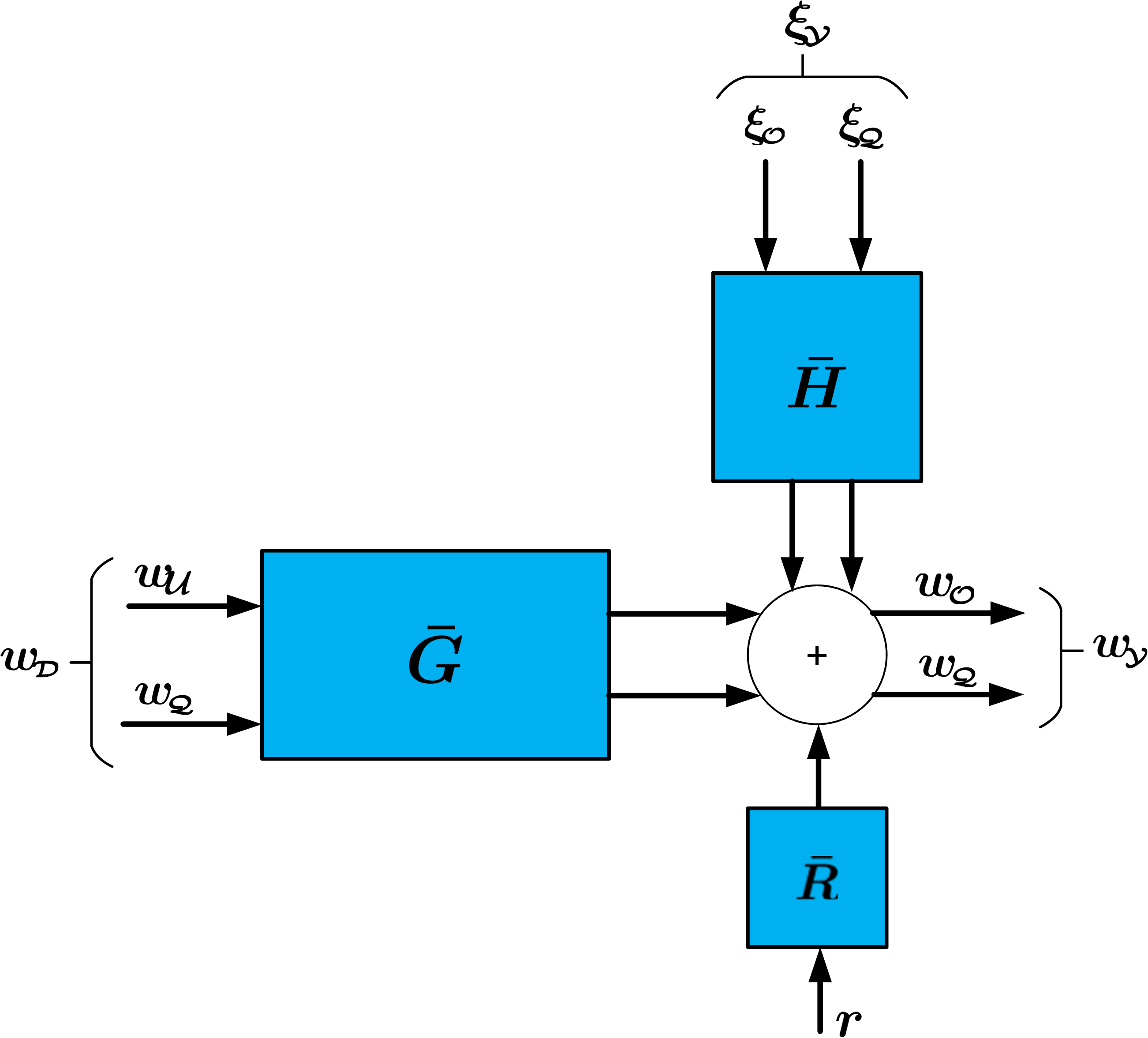}
	\caption{Figure to depict the classification of different sets of signals in the input and output of the setup.}
	\label{fig:Idsetup}
\end{figure}

We will denote $w_{\smY}$ as the node signals in $w$ that serve as predicted outputs with $w_j \in w_{\smY}$, and $w_{\smD}$ as the node signals in $w$ that serve as predictor inputs with $w_i \in w_{\smD}$. Next we decompose $w_{\smY}$ and $w_{\smD}$ into disjoint sets according to: $ \mY  =  \mQ \cup \mO \ ; \ \mD  =  \mQ \cup \mU$ where $w_{\smQ}$ are the node signals that are common in $w_{\smY}$ and $w_{\smD}$; $w_{\smO}$ are the node signals that are only in $w_{\smY}$; $w_{\smU}$ are the node signals that are only in $w_{\smD}$. A pictorial representation of the setup with classification of different sets of signals is presented in figure \ref{fig:Idsetup}. The remaining node signals in the network are $w_{\smZ}$ with $\mZ = \mL \backslash \mW$, where $\mL = \{1,2,\cdots, L\}$ and $\mW = \mD \cup \mY = \mU \cup \mY$. We denote $\bar\mY^{(j)} = \mY\backslash\{j\}$. Let $w_\ell, \ell \in \mD_j^w \subseteq \mW \backslash \{j\}$ denote the node signals in $w$ that have unmeasured paths\footnote{An unmeasured path is a path that does not pass through a node $w_{\ell}, \ell \in \mW$. Analogously, we can define unmeasured loops through a node $w_k$.} to $w_j$ and $u_{\ell}, \ell \in \mD_j^u$ denote the non-zero excitation signals in $u$ that have unmeasured paths to $w_j$. Also, for all $k \in \bar\mY^{(j)}$, let $\mD_k^w = \mW\backslash\{k\}$ and let $u_{\ell}, \ell \in \mD_k^u \subseteq \mW\cup\mZ$ denote the non-zero excitation signals in $u$.

\section{An empirical Bayes identification technique with missing node observations}\label{sec:EB}
% Consider a dynamic network defined by \eqref{eq.dgsMatrix}, however with $cov(e) = I$ and $H^0$ not necessarily monic (we refer to this network as the \emph{original network} in this paper).

%\vspace*{-8mm}
\subsection{Introduction}
We first write a MIMO representation of (part of) our network with inputs $(w_{\smQ},w_{\smU},u)$ and outputs $(w_{\smQ},w_{\smO})$. This is realized by the following steps\footnote{From now on, $^0$ is dropped for convenience.}:
\begin{enumerate}
	\item Firstly, similar to the reasoing in \cite{Ramaswamy&etal_TAC:21}, we write the system's equations for the measured variables as
	\beq
	\label{eq5a_4}
	\!\!\underbrace{\begin{bmatrix} w_{\smQ} \\ w_{\smO} \\ \hline w_{\smU} \end{bmatrix}}_{w_m}\!\! =\!\!
	\underbrace{\left[ \begin{array}{c|c} \!\bar G & 0 \!\\ \hline \\[-3mm]\! \bar G_{\smU\smD} & \bar G_{\smU \smO} \!\end{array} \right]}_{\bar G_m}\!
	\!\begin{bmatrix} w_{\smQ} \\ w_{\smU} \\ \hline w_{\smO} \end{bmatrix}\! +\!
	\underbrace{\left[ \begin{array}{c|c} \bar H & 0 \!\\ \hline \\[-3mm] 0 \ & \bar H_{\smU\smU}\! \end{array} \right]}_{\bar H_m}\!
	%\begin{bmatrix} \bar H_{11}\! &\! |\! &\! 0 \\ \hline \bar H_{21}\! &\! |\! &\! \bar H_{22} \end{bmatrix}
	\!\underbrace{\begin{bmatrix} \xi_{\smQ} \\ \xi_{\smO} \\ \hline \xi_{\smU} \end{bmatrix}}_{\xi_m}\!\! + \!\!
	\underbrace{\left[ \begin{array}{c} \!\bar R \!\\ \hline \\[-3mm] \!\bar R_{\smU}\! \end{array} \right]}_{\bar R_m}\!\! u
	\eeq
	with $\xi_m$ a white noise process, while $\bar H$ is monic, stable and stably invertible and the components in $\bar G$ are zero if it concerns a mapping between identical signals. This step is made by removing the non-measured signals $w_{\smZ}$ from the network, while maintaining the second order properties of the remaining signals. This step is referred to as immersion of the nodes in $w_{\smZ}$ \cite{Dankers&etal_TAC:16}. After immersion, we re-write the system's equation to structure the noise model such that there are no confounding variable between $w_{\smU} \leftrightarrow w_{\smY}$ given $w_{\smW}$.
	\item As an immediate result of the previous step we can write an expression for the output variables $w_{\smY}$, by considering the upper part of the equation (\ref{eq5a_4}), as
	\beq
	\label{eq5_4}
	\underbrace{\begin{bmatrix} \!w_{\smQ}\! \\ \!w_{\smO}\! \end{bmatrix}}_{w_{\smY}}\!\! =\!\!
	\underbrace{\begin{bmatrix} \!\bar G_{\smQ\smQ}\! &\! \bar G_{\smQ\smU}\!  \\
			\!\bar G_{\smO\smQ}\! & \!\bar G_{\smO\smU}\! \end{bmatrix}}_{\bar G}
	\underbrace{\begin{bmatrix} \!w_{\smQ}\! \\ \!w_{\smU}\! \end{bmatrix}}_{w_{\smD}} +
	\underbrace{\begin{bmatrix} \bar H_{\smQ\smQ} & \bar H_{\smQ \smO} \\
			\bar H_{\smO\smQ} & \bar H_{\smO\smO} \end{bmatrix}}_{\bar H}
	\underbrace{\begin{bmatrix} \xi_{\smQ} \\ \xi_{\smO}  \end{bmatrix}}_{\xi_{\smY}} + \bar R u
	\eeq
	with $cov(\xi_{\smY}) := \bar\Lambda$.
\end{enumerate}

Now we will show that we can write an expression for the output variables $w_{\smY}$ as in \eqref{eq5_4}.
\begin{proposition}\label{prop3a}
	Consider a dynamic network defined by \eqref{eq.dgsMatrix}, however with $cov(e) = I$ and $H^0$ not necessarily monic. Then,
	\begin{enumerate}
		\item
		there exists a representation (\ref{eq5a_4}) of the measured node signals $w_m$, with $\bar H_m$ monic, stable and stably invertible, and $\xi_m$ a white noise process, and
		\item for this representation there are no
		confounding variable between $w_{\smU} \leftrightarrow w_{\smY}$ given $w_{\smW}$.
	\end{enumerate}
%	The systems equations for the output variables in $w_{\smY}$ can always be written as,
%	\vspace{-0.15cm}
%	\beqr
%	\label{eq5_4}
%	w_{\smY} = \bar Gw_{\smD} + \bar H\xi_{\smY} + \bar Ru,
%	\eeqr
%	\noindent where $\xi_{\smY}$ a white noise process with dimensions conforming to $w_{\smY}$, with $cov(\xi_{\smY}) = \bar\Lambda$ and with $\bar H$ being monic, stable and stably invertible.
	%Such a representation can always be found.
\end{proposition}
{\bf Proof:} See Proposition 4.2 in \cite{Ramaswamy_diss}.

The consequence of Proposition \ref{prop3a} is that the output node signals in $w_{\smY}$ can be explicitly written in the form of (\ref{eq5_4}), in terms of input node signals $w_{\smD}$, excitation signals $u$ and disturbances, without relying on node signals in $w_{\smZ}$.
% For the representation \eqref{eq5_4}, the structure of $\bar R$ will be induced by the topological properties of the network. Note that, $\bar Ru$ does not necessarily include all external excitation signals in the network, because the effect of some of them on $w_{\smY}$ will be incorporated in the term $\bar Gw_{\smD}$. This latter group of external signals will be referred to as $u_{\bar\smP}$, with $\bar \mP$ to be defined later on. In order to use \eqref{eq5_4} for identifying $\bar G$ it is attractive to further explore the structure of $\bar R$, i.e. to determine which elements of $\bar R$ are fixed (e.g. 1 or 0) and which terms are dynamic. To this end, we present the Lemma \ref{prop43}.
From the result of Proposition \ref{prop3a}, it is known that the system's equations for the output variables $w_{\smY}$ can be written as,
\beq\label{eq:10.1}
w_{\smY} = \bar Gw_{\smD} + \bar H \xi_{\smY} + \bar Ru
\eeq
with $\xi_{\smY}$ a Gaussian white noise process, while $\bar H$ is full matrix, monic, stable and stably invertible.

For estimation purposes we are going to use a specific form of \eqref{eq:10.1}, as formulated next.
\begin{condition}\label{condx2ab}
	There are no confounding variable between $w_{j} \leftrightarrow w_{\smW\backslash\{j\}}$ given $w_{\smW}$.
\end{condition}
\begin{condition}\label{cond:path_OtoJ}
	All paths from $w_{h}, h \in \mO\backslash\{j\}$ to $w_{j}$ pass through a node in $w_{\smW}$.
\end{condition}
\begin{proposition}\label{prop:10.1}
	Consider the sitauation of Proposition \ref{prop3a}. If Conditions \ref{condx2ab} and \ref{cond:path_OtoJ} are satisfied, then there exists a form \eqref{eq:10.1} with $\bar H$ being (block) lower triangular as follows,
	\beq
	\label{eq51}
	\underbrace{\begin{bmatrix} \!w_{j}\! \\ \!w_{\bar\smY^{(j)}}\! \end{bmatrix}}_{w_{\smY}}\!\! =\!\!
	\underbrace{\begin{bmatrix} \!\bar G_{j\smD}\!  \\
			\!\bar G_{\bar\smY^{(j)}\smD}\! \end{bmatrix}}_{\bar G}
	\!w_{\smD} \!+\!
	\underbrace{\begin{bmatrix} \bar H_{jj} & 0 \\
			\bar H_{\bar\smY^{(j)} j} & \bar H_{\bar\smY^{(j)}\bar\smY^{(j)}} \end{bmatrix}}_{\bar H}\!
	\underbrace{\begin{bmatrix} \xi_{j} \\ \xi_{\bar\smY^{(j)}} \end{bmatrix}}_{\xi_{\smY}} \!+ \bar R u.
	\eeq
\end{proposition}
\vspace{-10pt}
{\bf Proof:} See Proposition 10.1 in \cite{Ramaswamy_diss}.

\vspace{-10pt}
\begin{condition}
	\label{cond5}
	Let $G_{ji}$ be the target network module to be identified.
	In the network (\ref{eq.dgsMatrix}):
	\begin{itemize}
		\item Every path from $w_i$ to $w_j$, excluding the path through $G_{ji}$, passes through a node $w_k, k \in \mW$, and
		\item Every loop through $w_j$ passes through a node in $w_k, k \in \mW$. \hfill $\Box$
	\end{itemize}
\end{condition}
\vspace{-10pt}
In order to guarantee that $\bar G_{ji}(q) = G_{ji}(q)$, i.e the target module appearing in equation (\ref{eq51}) is the target module of the original network \eqref{eq.dgsMatrix} \emph{(invariance of target module)}, we utilize the following result:
\begin{proposition}\label{prop:10.2}
	Consider the situation of Proposition \ref{prop:10.1}. If condition \ref{cond5} is also satisfied, then we have $\bar G_{ji}(q) = G_{ji}(q)$.
\end{proposition}
{\bf Proof:} See Proposition 10.2 in \cite{Ramaswamy_diss}.

\vspace{-10pt}
Based on the above results, we can now re-write \eqref{eq:10.1} in the predictor form as given below.
\vspace{-10pt}
\begin{proposition}\label{prop1}
	Consider the network as represented by \eqref{eq.dgsMatrix} where the set of all nodes $w_{\smL}$ is decomposed into disjoint sets $w_{\smQ}$, $w_{\smO}$, $w_{\smU}$ and $w_{\smZ}$ such that Conditions \ref{condx2ab}, \ref{cond:path_OtoJ} and \ref{cond5} are satisfied. The network equation for the node $w_j$ and $w_k, k \in \bar\mY^{(j)}$ can be written as\footnote{from now on superscript $^0$ is dropped for convenience.},
	\begin{align}
		w_j(t) &= \hat w_j(t|t-1) + \xi_j(t) \nonumber \\
		&= {S_j(q)}(w_j(t)-u_j(t)) \!+\! (1-S_j(q))G_{ji}(q)w_i(t) \nonumber\\
		&+\!\!\!\! \sum_{\ell \in \mathcal{D}_j^w\backslash \{i\}}\!\!\!\!S_{\ell}(q)w_{\ell}(t) \!+\! u_j(t) \!+\!\!\! \sum_{\ell \in \mD_j^u \backslash \{j\}}\!\! S_{j\ell}(q)u_{\ell}(t) \!+\! \xi_j(t),\label{eq:outputeq1}\\
		w_k(t) = &\hat w_k(t|t-1) + \xi_k(t) \nonumber\\
		&= {B_k(q)}(w_k(t) - u_k(t)) + \sum_{\ell \in \mD_k^w} {B_\ell(q)}w_\ell(t) \nonumber\\
		&+ \sum_{\ell \in \mD_k^u} B_{k\ell}(q)u_\ell(t) + u_k(t) + \xi_k(t), \label{eq:missingeq1}
		% 	&w_a(t) = {F_a(q)}(w_a(t) - r_a(t)) + \sum_{\ell \in \mD_a^w}{F_\ell(q)}w_\ell(t) \nonumber\\
		% 	&+ \sum_{\ell \in \mD_a^r} F_{a\ell}(q)r_\ell(t) + r_a(t) + \xi_a(t),\label{eq:additionalnodeq1}
	\end{align}
	where we isolate the target module $G_{ji}(q)$. $S_{\star}(q), B_{\star}(q)$ are strictly proper predictor filters, and $\xi_{\smY}(t)$ is Gaussian white noise with $cov(\xi_{\smY}) = \bar\Sigma$.
	% 	$$
	% 	cov(\begin{bmatrix}
	% 		\xi_{j} \\ \xi_{m} \\ \xi_a
	% 	\end{bmatrix}) := \Sigma = \begin{bmatrix}
	% 		\sigma_{j}^2 & \sigma_{jm}^2 & \sigma_{ja}^2 \\
	% 		\sigma_{jm}^2 & \sigma_{m}^2 & \sigma_{ma}^2 \\
	% 		\sigma_{ja}^2 & \sigma_{ma}^2 & \sigma_{a}^2 \\
	% 	\end{bmatrix}.
	% $$
\end{proposition}
\vspace{-10pt}
{\bf Proof:} See Proposition 10.3 in \cite{Ramaswamy_diss}.

The lower (block) block triangular structure of $\bar H$ in \eqref{eq51} allows us to isolate the target module $G_{ji}$ in \eqref{eq:outputeq1}. Realizing the above representation requires conditions on the selection of node signals in $w_{\smQ}, w_{\smO},w_{\smU}$ (i.e. conditions \ref{condx2ab}, \ref{cond:path_OtoJ} and \ref{cond5}). The conditions are not a restriction. They can always be satisfied by appropriate selection of signals in the sets $\mQ, \mO$ and $\mU$.

\subsection{Predictor model}
The identification method developed in this paper relies on data augmentation strategies \cite{Tanner&Wong_ASA:1987, Vandyk&Meng_CGS:2001}. Using this strategy, we treat a non-measured missing node signal as a latent variable which is estimated along with the parameters, while adding measured node signals that are ascendants and descendents of the missing node to our estimation problem. In this way signals that carry information on the missing node signal are incorporated in our estimation procedure. For simplicity, we are going to assume one missing node $w_m$ (i.e. the cardinality of $\mZ_j$ is one). The procedure can simply be generalized to multiple missing node signals.

We are now going to construct a predictor model with inputs $w_{\smD}$ and outputs $w_{\smY}$ according to the following procedure:

\begin{enumerate}
    \item Select $w_{\smD}$ to be a set of signals, including $w_i$ and $w_m$, that satisfies the parallel path and loop condition \ref{cond5}.
    \item Select $w_{\smY}$ to include $w_j$, while satisfying Conditions \ref{condx2ab} and \ref{cond:path_OtoJ}. Include $w_m$ in $w_{\smY}$.
    \label{item2}
    \item Select all measured descendant nodes $w_a$ that have an unmeasured path from $w_m$, and include them in the output $w_{\smY}$.
    \item Select all measured ascendant nodes that have an unmeasured path to $w_m$, and include them in the input $w_{\smD}$.
    \item Select all measured ascendant nodes that have an unmeasured path to $w_a$, and include them in the input $w_{\smD}$.
%
%    measured node $w_a$ that is a descendant node of $w_m$, other than %$w_j$, that has an unmeasured path from $w_m$, and add this node to %$w_{\smY}$.
%    \item Find an ascendant node of $w_m$, i.e. a node that reaches %$w_m$ through an unmeasured path, and add this node to $w_{\smD}$. %\red{[This step is not needed. This node is already covered in Step 6.]}
%    \item Include all measured signals in $w_{\smD}$ that have %unmeasured paths to a node in $w_{\smY}$.
\end{enumerate}
Note that all signals in $w_{\smD} \cup w_{\smY}$ are measured node signals, except for the missing node $w_m$. However, for all considerations on confounding variables and unmeasured paths, the missing node signal $w_m$ is considered to be a measured node signal.

%\blue{In this procedure the information of the missing node $w_m$ is %replaced by an ascendant node to be included in $w_{\smD}$ and a %descendant node, to be included in $w_{\smY}$.}

\begin{remark}
Satisfying Conditions \ref{condx2ab} and \ref{cond:path_OtoJ} in Step \ref{item2} might not be feasible based on measured signals only. In that case the blocking of paths, e.g. the blocking of confounding variables, can be realized by introducing additional missing nodes, for which the same stepwise procedure above is followed.
\end{remark}

%However, the method described in this chapter can be extended to any number of missing nodes and additional nodes. As mentioned above, we are going to formulate
The result of the above procedure is
a MIMO model structure where the outputs are (1) the output of the target module i.e. $w_j$, (2) the missing node output $w_m$, (3) additional nodes $w_a$ that are descendants of the missing node, leadng to $\mY = \{w_j, w_m, w_a\}$. The inputs of the predictor model are collected in $w_{\smD}$ which includes $w_m$.
%consists of any set of measurable node signals, and has to contain the missing node signal(s) $w_m$. We include the measurable signals that have unmeasured paths to $w_{\smY}$ in $w_{\smD}$ along with the missing node signal(s). The addition of missing node signal(s) in $w_{\smW}$ ensures that condition \ref{cond5} is satisfied. Also, as discussed earlier, we can choose signals in $w_{\smY}$ and $w_{\smD}$ such that conditions \ref{condx2ab} and \ref{cond:path_OtoJ} are satisfied.
Since $w_m$ and $w_a$ are both in the set $\bar \mY^{(j)}$ they can be written according to \eqref{eq:missingeq1} as,
%to As a result of the proposition \ref{prop1}, we can write the equation for the missing node $w_m$ and additional node $w_a$ as,
\begin{align}
	w_m(t) \!\!&=\!\! {B_m(q)}(w_m(t) \!-\! u_m(t)) \!+\! \sum_{\ell \in \mD_m^w} {B_\ell(q)}w_\ell(t) \nonumber\\
	&\!+\! \sum_{\ell \in \mD_m^u} B_{m\ell}(q)u_\ell(t) \!+\! u_m(t) \!+\! \xi_m(t),\label{eq:missingnodeq1}
	\end{align}
	\begin{align}
	w_a(t) \!\!&=\!\! {F_a(q)}(w_a(t) \!-\! u_a(t)) \!+\! \sum_{\ell \in \mD_a^w}{F_\ell(q)}w_\ell(t) \nonumber\\
	&\!+\! \sum_{\ell \in \mD_a^u} F_{a\ell}(q)u_\ell(t) \!+\! u_a(t) \!+\! \xi_a(t),\label{eq:additionalnodeq1}
\end{align}
with $cov(\begin{bmatrix}
		\xi_{j}(t) \\ \xi_a(t) \\ \xi_{m}(t)
	\end{bmatrix}) := \bar\Sigma = \begin{bmatrix}
		\sigma_{j}^2 & 0 & 0 \\
		0 & \sigma_{a}^2 & \sigma_{am}^2 \\
		0 & \sigma_{am}^2 & \sigma_{m}^2
	\end{bmatrix}$,
where for simplicity of notation we restrict to the situation where both $w_m$ and $w_a$ are scalar. This can very easily be generalized to the situation of multivariate $w_m$ and/or $w_a$.

From the above, it is very clear that if we use the additional node in the output, we need to model additional modules. This increase in complexity counterbalances the gain obtained by using more information. In this paper, we develop an identification framework that uses the additional node(s); we also provide the framework that does not use additional nodes as a special case of the former.

\subsection{Vector description of the dynamics}
In this section, we obtain a vector description of the network dynamics for the available $N$ measurements. For notation purposes, we introduce the $N$-dimensional vector $g_{ji}$ (which will also depend on $\theta$, although we will keep this dependence tacit) as the first $N$ coefficients of the impulse response of $G_{ji}(q, \theta)$. Similarly, we define the vector ${s_{k}}$, $k \in \{\mathcal D_j^w\} \backslash i$, ${s_{jk}}$, $k \in \{\mathcal D_j^u\} \backslash j$ and $s_j$ as the vectors containing the first $l$ coefficients of the impulse responses of ${S_{k}}(q)$, $k \in \{\mathcal D_j^w\} \backslash i$, ${S_{jk}}(q)$, $k \in \{\mathcal D_j^u\} \backslash i$, and $S_j(q)$, respectively. Similarly, $b_k$, $b_{mk}$, $b_m$, $f_k$, $f_{ak}$, $f_a$ are defined as the vectors containing the first $l$ coefficients of the impulse responses of $B_{k}$, $B_{mk}$, $B_{m}$, $F_{k}$, $F_{ak}$, $F_{a}$ respectively. The integer $l$ is chosen large enough to ensure $s_{k}(l+1), s_{jk}(l+1), s_{j}(l+1), b_{k}(l+1), b_{mk}(l+1), b_{m}(l+1), f_{k}(l+1), f_{ak}(l+1), f_{m}(l+1) \simeq 0$.

\begin{lemma}\label{lemma2}
	Let the vector notation for the node $w_{k}(t)$ be $w_{k} := \begin{bmatrix}
	w_{k}(1) & \ldots & w_{k}(N)
	\end{bmatrix}^T$ where $k \in \{j, m, a\}$. Considering the parameterization of $G_{ji}(q)$ (i.e. $G_{ji}(q,\theta)$), the network dynamics in \eqref{eq:outputeq1}, \eqref{eq:missingnodeq1} and \eqref{eq:additionalnodeq1} can be represented in the vector form as:
	\beqr\label{eq:vectorwm}
	w_m \!\!& = &\!\! \tilde W_mb_m + \sum_{k \in \mD_m^w} W_{k} b_{k} +\sum_{k \in \mD_m^u} R_{k} b_{mk} + u_m + \xi_m,
	\eeqr
	\beqr
	w_j \!\!& = &\!\! \tilde W_j s_j \!+\! W_{ji}g_{ji} \!+\!\!\!\!\!\! \sum_{k \in \mathcal{D}_j^w \backslash \{i \}} \!\!\!\!\!\!W_{k}s_{k} +\!\!\!\!\!\!\sum_{k \in \mathcal{D}_j^u \backslash \{j \}} \!\!\!\!\!\!R_{k}s_{jk} \!+\! u_j \!+\! \xi_j,\label{eq:vectorwj} \\
	w_a \!\!& = &\!\! \tilde W_af_a + \sum_{k \in \mD_a^w} W_{k} f_{k} + \sum_{k \in \mD_a^u} R_{k} f_{mk} + u_a + \xi_a,\label{eq:vectorwa}
	\eeqr
	where $\xi_j$, $\xi_m$, $\xi_a$ are the vectorized noise and $r_j$, $r_m$, $r_a$ are the vectorized excitation signal. $\tilde W_j$, $\tilde W_m$, $\tilde W_a$, $W_{ji}$, $W_k$ and $R_k$ are Toeplitz matrices constructed from measurements of the respective node and excitation signals.
\end{lemma}
{\bf Proof:}
	We denote by $W_k \in \mathbb{R}^{N\times l}$ the Toeplitz matrix of the vector $\overrightarrow w_k := \begin{bmatrix}
	0 & w_k(1) & \ldots & w_k(N-1) \end{bmatrix}^T$, $k \in \mathcal{D}_j^w \cup \mD_m^w \cup \mD_a^w \cup \mY$ and $W_{ji} \in \mathbb{R}^{N\times N}$ the Toeplitz matrix of the vector $\overrightarrow w_i := \begin{bmatrix}
	0 & w_i(1) & \ldots & w_i(N-1) \end{bmatrix}^T$.  Let $R_{\ell} \in \mathbb{R}^{N\times l}$ be the Toeplitz matrix of the vector $\overrightarrow u_{\ell} := \begin{bmatrix}
	0 & r_{\ell}(1) & \ldots & r_{\ell}(N-1) \end{bmatrix}^T$ where $\ell \in \mY \cup \mD_m^u \cup \mD_a^u \cup \mD_j^u$. Similarly, we denote by $\overleftrightarrow W_k \in \mathbb{R}^{N\times l}$ the Toeplitz matrix of the vector $\overleftrightarrow{w}_k := \begin{bmatrix} 0 & 0 & -w_k(1) & \ldots & -w_k(N-2) \end{bmatrix}^T$, $k \in \{i,j\}$, and by {$G_{\theta}$} the Toeplitz of $g_{ji}$. Considering the parameterization of $G_{ji}^0$ and the above established notations, we can rewrite the network dynamics in \eqref{eq:outputeq1} as \eqref{eq:vectorwj}, \eqref{eq:additionalnodeq1} as \eqref{eq:vectorwa}, and \eqref{eq:missingnodeq1} as \eqref{eq:vectorwm} where $\tilde W_j := W_j - R_j + {G_{\theta}}\overleftrightarrow W_{i}$, $\tilde W_m := W_m - R_m$, $\tilde W_a := W_a - R_a$.

%\begin{corollary}
%	Consider the situation in Lemma \ref{lemma2}. When all paths through the node $w_m$ in the original network \eqref{eq.dgsMatrix} passes through nodes in $w_{\smD_m}$, then the network dynamics in \eqref{eq:missingnodeq} can be represented in the vector form as:
%	\beqr
%	w_m \!=\! \tilde W_mp_m \!+\! \sum_{k \in \mathcal{D}_m} \!\!W_{k} p_{mk} \!+\!\!\!\!\! \sum_{k \in \mU_m^r \backslash \{m\}} \!\!\!\!\!\!R_{k} p_{mk} \!+\! r_m \!+\! \xi_m.
%	\eeqr
%
%{\bf Proof:} The proof directly follows from the result of Lemma \ref{lemma 1} and the reasoning of Lemma \ref{lemma2}.
%\end{corollary}
%We next discuss the strategy to avoid parameterization of the additional modules (all modules except the target module) in the MIMO structure using regularized kernel-based methods  and also how we tackle the problem of missing node observations in the direct method framework.

\subsection{Strategy to reduce the number of parameters for nuisance modules}
Before we move to an estimation scheme that can deal with the missing node $w_m$, we explain how we can avoid having to estimate a huge number of parameters in the nuisance modules, i.e. the modules that we need to identify but that are not the target module of interest. For this we follow the work in \cite{Ramaswamy&etal_Autom:2021}.
%Having represented the network dynamics in vector form as in \eqref{eq:vectorwm}, \eqref{eq:vectorwj} and \eqref{eq:vectorwa}, and parameterized the target module as $G(q,\theta)$, we now discuss the modeling strategy of the nuisance modules (i.e. modules other than the target module).
%The approach described in this section is an extension of our previous work
where no missing nodes are assumed and extend it to the situation where there are missing node(s). Our goal is to limit the number of parameters necessary to describe $w_j$, $w_a$ and $w_m$, in order to increase the accuracy of the estimated parameter vector of interest $\theta$. Therefore, while we keep a parametric model for $G_{ji}$, for the remaining impulse responses in \eqref{eq:vectorwm}, \eqref{eq:vectorwj} and \eqref{eq:vectorwa}, we use nonparametric models induced by Gaussian processes \cite{Rasmussen2006book}. The choice of Gaussian processes is motivated by the fact that, with a suitable choice of the prior covariance matrix, we can get a {significant} reduction in the variance of the estimated impulse responses \cite{Pillonettoetal_Autom:14}.  Therefore, we model $s_k, k \in j \cup \mD_j^w\backslash\{i\}$, $s_{jk}$, $k \in \mathcal{D}_j^u$, $b_k, k \in m \cup \mD_m^w$, $b_{mk}$, $k \in \mathcal{D}_m^u$, $f_k, k \in a \cup \mD_a^w$, $f_{ak}$, $k \in \mD_a^u$ as independent Gaussian processes (vectors in this case) with zero-mean. The covariance matrix of these vectors, usually referred to as a kernel in this context, is chosen to be corresponding to the so-called \textit{First-order Stable Spline kernel}.
%\footnote{It is clear that these impulse responses share some common dynamics given by the pre-multiplication with the inverse of the noise model $H_j(q)$. However, for computational purposes it is convenient to treat the impulse responses as independent. Furthermore, incorporating the mutual dependence through a suitable choice of prior distribution seems a non-trivial problem that deserves a thorough analysis that is outside the scope of this chapter.}
The general structure of this kernel is given by
% \begin{equation} \label{eq:ssk}
$
\lambda[{K_\beta}]_{x,y} = \lambda\beta^{\max(x,y)} \,,
% \end{equation}
$
where ${\beta} \in [0, 1)$ is a \emph{hyperparameter} that regulates the decay velocity of the realizations of the corresponding Gaussian vector, while $\lambda \geq 0$ tunes their amplitude. The choice of this kernel is motivated by the fact that it enforces favorable properties such as stability and smoothness in the estimated impulse responses \cite{Pillonetto&DeNicolao_Autom:11}, \cite{Pillonetto&DeNicolao_Autom:10}. Therefore, we have that
\begin{align}
{s_{k}} & \sim \mathcal{N}(0, \lambda_k^s K_{\beta_k^s})\quad ,\, k \in j \cup \mathcal{D}_j^w \backslash \{i \}, \label{eq:prior_sj} \\
{s_{jk}} & \sim \mathcal{N}(0, \lambda_{jk}^s K_{\beta_{jk}^s}) \quad,\, k \in \mathcal{D}_j^u \backslash \{j \}, \label{eq:prior_sk} \\
{b_{k}} & \sim \mathcal{N}(0, \lambda_k^b K_{\beta_k^b})\quad,\, k \in m \cup \mathcal{D}_m^w, \label{eq:prior_pj} \\
{b_{mk}} & \sim \mathcal{N}(0, \lambda_{mk}^b K_{\beta_{mk}^b}) \quad,\, k \in \mathcal{D}_m^u, \label{eq:prior_pk} \\
{f_{k}} & \sim \mathcal{N}(0, \lambda_k^f K_{\beta_k^f})\quad,\, k \in a \cup \mathcal{D}_a^w, \label{eq:prior_fj} \\
{f_{ak}} & \sim \mathcal{N}(0, \lambda_{ak}^f K_{\beta_{ak}^f}) \quad,\, k \in \mD_a^u, \label{eq:prior_fk}
\end{align}
where we have assigned different hyperparameters to the impulse response priors to guarantee flexible enough models. In \eqref{eq:vectorwm} - \eqref{eq:vectorwa} we have terms that are multiplication of the Toeplitz matrix related to missing node $w_m$ and the impulse response models, where both the missing node data and the prior hyperparameters of the impulse models are unknown and need to be estimated. In order to tackle the identifiability issues, we set the hyperparameters $\lambda_m^s$, $\lambda_{m}^b$ and $\lambda_{m}^f$ (i.e. $\lambda$'s corresponding to the modules having missing node as inputs) to be 1.

\subsection{Incorporating Empirical Bayes approach}
We now explain how the parameters of the priors and the target module are estimated using the Empirical Bayes approach. For this, we define
\begin{equation}
s := \begin{bmatrix}
s_j^\top & s_{c_1}^\top & \dots & s_{c_p}^\top  & {s_{jk}}_1^\top & {s_{jk}}_2^\top & \dots & {s_{jk}}_p^\top
\end{bmatrix}^\top \,,
\end{equation}
where $c_1,\,\ldots,\,c_p$ and $k_1,\,\ldots,\,k_p$ are the elements of the set $\mD_j^w \backslash \{i\}$ and $\mD_j^u \backslash \{j\}$, and
\begin{equation}
\mathbf W := \begin{bmatrix}
\tilde W_j & {W_{c_1}} & {W_{c_2}} & \dots & {R_{k_{p-1}}} & {R_{k_p}}
\end{bmatrix}\,,
\end{equation}
\begin{equation}
\mathbf K_{1} := \mathrm{diag}\lbrace {\lambda_j^s}{K_{\beta_j^s}}, {\lambda_{c_1}^s}{K_{{\beta_{c_1}^s}}}, \dots, {\lambda_{jk_p}^s}{K_{{\beta_{jk_p}^s}}}\rbrace .
\end{equation}
Analogously we define
\begin{equation}
b := \begin{bmatrix}
b_m^\top  & b_{c_1}^\top & \dots & b_{c_p}^\top  & {b_{mk}}_1^\top & {b_{mk}}_2^\top & \dots & {b_{mk}}_p^\top
\end{bmatrix}^\top \,,
\end{equation}
\begin{equation}
\mathbf R := \begin{bmatrix}
\tilde W_m & {W_{c_1}} & {W_{c_2}} & \dots & {R_{k_{p-1}}} & {R_{k_p}}
\end{bmatrix}\,,
\end{equation}
\begin{equation}
\mathbf K_{2} := \mathrm{diag}\lbrace {K_{\beta_m^b}}, {\lambda_{c_1}^s}{K_{{\beta_{c_1}^b}}}, \dots, {\lambda_{mk_p}^b}{K_{{\beta_{mk_p}^b}}}\rbrace.
\end{equation}
where $c_1,\,\ldots,\,c_p$ and $k_1,\,\ldots,\,k_p$ are the elements of the set $\mathcal{D}_m^w$ and $\mD_m^u$ respectively, and
\begin{equation}
f := \begin{bmatrix}
f_a^\top & f_{c_1}^\top & \dots & f_{c_p}^\top & {f_{ak}}_1^\top & {f_{ak}}_2^\top & \dots & {f_{ak}}_p^\top
\end{bmatrix}^\top \,,
\end{equation}
\begin{equation}
\mathbf Q := \begin{bmatrix}
\tilde W_a & {W_{c_1}} & {W_{c_2}} & \dots & {R_{k_{p-1}}} & {R_{k_p}}
\end{bmatrix}\,,
\end{equation}
\begin{equation}
\mathbf K_{3} := \mathrm{diag}\lbrace {\lambda_{a}^f}{K_{\beta_a^f}}, {\lambda_{c_1}^f}{K_{{\beta_{c_1}^f}}}, \dots, {\lambda_{ak_p}^f}{K_{{\beta_{ak_p}^f}}}\rbrace.
\end{equation}
where $c_1,\,\ldots,\,c_p$ and $k_1,\,\ldots,\,k_p$ are the elements of the set $\mathcal{D}_a^w$ and $\mD_a^u$ respectively.
Using the above, we can rewrite \eqref{eq:vectorwm}, \eqref{eq:vectorwj} and \eqref{eq:vectorwa} in compact form and we obtain the following model:
\beqr\label{eq:model}
%w_m & = & \mathbf R b + r_m + \xi_m \, , \nonumber\\
\underbrace{\begin{bmatrix}
		w_{j} \\ w_a \\ w_m
\end{bmatrix}}_{w_{\smY}} \!\!&=&\!\! \underbrace{\begin{bmatrix}
		\mathbf W & \mathbf 0 & \mathbf 0 \\ \mathbf 0 & \mathbf Q & \mathbf 0 \\ \mathbf 0 & \mathbf 0 & \mathbf R
\end{bmatrix}}_{\mathbf W_{\smD}}\underbrace{\begin{bmatrix}
		s \\ f \\ b
\end{bmatrix}}_{g} \!\!+\!\! \underbrace{\begin{bmatrix}
		W_{ji} \\ \mathbf 0 \\ \mathbf 0
\end{bmatrix}}_{\mathbf W_{ji}}g_{ji} \!+\! \underbrace{\begin{bmatrix}
		u_{j} \\ u_a \\ u_m
\end{bmatrix}}_{u_{\smY}}  \!\!+\!\! \underbrace{\begin{bmatrix}
		\xi_j \\ \xi_a \\ \xi_m
\end{bmatrix}}_{\xi_{\smY}} \, , \nonumber\\
s & \sim & \mathcal{N}(0, \mathbf K_1) \, , \nonumber\\
b & \sim & \mathcal{N}(0, \mathbf K_2) \, ,\\
f & \sim & \mathcal{N}(0, \mathbf K_3) \, , \nonumber\\
\xi_{\smY} & \sim & \mathcal{N}(0, \Sigma) \, , \nonumber
%\underbrace{\begin{bmatrix}
%	\xi_j \\ \xi_a \\ \xi_m
%\end{bmatrix}}_{\xi_{\bar\smY}} & \sim & \mathcal{N}\left(\mathbf 0, \underbrace{\begin{bmatrix}
%\sigma_j^2 I & \mathbf 0 & \mathbf 0 \\
%\mathbf 0 & \sigma_{a}^2 I & \sigma_{am}^2 I\\
%\mathbf 0 & \sigma_{am}^2 I & \sigma_{m}^2 I
%\end{bmatrix}}_{\Sigma}\right), \nonumber
\eeqr
where $s, b, f$ and $\xi_{\smY}$ are mutually independent and with $\Sigma = \bar\Sigma \otimes I_N$.
%\beqr\label{eq:model}
%w_m & = & \mathbf R p + r_m + x_m \, , \nonumber\\
%w_j & = & \mathbf W s + W_{ji}g_{ji} + \ell\left(w_m \!-\! \mathbf Rp \!-\! r_m \right) + r_j + x_j \, , \nonumber\\
%s & \sim & \mathcal{N}(0, \mathbf K_\beta) \, , \nonumber\\
%p & \sim & \mathcal{N}(0, \mathbf K_\gamma) \, \\
%x_j & \sim & \mathcal{N}(0, \sigma_j^2 I)\nonumber\\
%x_m & \sim & \mathcal{N}(0, \sigma_m^2 I) \nonumber
%\eeqr
%where $s, p, x_j, x_m$ are mutually independent.
%Having assumed a Gaussian distribution of the noise, we can write the joint probabilistic description of $s$ and $w_j$, which is jointly Gaussian, as:
%\begin{equation}
%p\Bigg(\begin{bmatrix}
%s \\ w_j
%\end{bmatrix}; \eta \Bigg) \sim \mathcal{N}\Bigg(\begin{bmatrix}
%0 \\ W_ig_{ji}
%\end{bmatrix}, \begin{bmatrix}
%\mathbf K & \mathbf K \mathbf W^\top\\ \mathbf W \mathbf K & \mathbf P
%\end{bmatrix}\Bigg),
%\end{equation}
%where
%\begin{equation}
%\mathbf P := \sigma_j^2I_N + \tilde W{\lambda_{j}}{{K_\beta}_j}\tilde W + \sum_{ k \in \mathcal{N}_j \backslash \{i \}} {W_{k}}{\lambda_{k}}{{K_\beta}_k}{W_{k}}^\top.
%\end{equation}
We note that the above model depends upon the vector of parameters
\beqr
&&\eta := [ \
\theta \
{\lambda_j^s} \
\dots \
{\lambda_{jk_p}^s} \
{\lambda_{c_1}^b} \
\dots \
{\lambda_{mk_p}^b} \ {\lambda_a^f} \
\dots \
{\lambda_{ak_p}^f} \nonumber\\
&&{\beta_j^s} \
\dots \
{\beta_{jk_p}^s} \
{\beta_m^b} \
\dots \
{\beta_{mk_p}^b} \ {\beta_a^f} \
\dots \
{\beta_{ak_p}^f} \
\sigma_j^2 \
\sigma_m^2 \
\sigma_a^2 \ \sigma_{am}^2 \ ] , \nonumber
\eeqr
which contains the parameter vector of the target module, the hyperparameters of the kernels of the impulse response models of the other modules, and the parameters related to the covariance of the noise corrupting $w_j(t)$, $w_a(t)$ and $w_m(t)$. Note that $\theta$ appears in $g_{ji}$ while the other parameters in $\eta$ appear in $g$ and in covariance of $\xi_{\smY}$. Therefore, we focus on the estimation of $\eta$, since it contains the parameter of interest $\theta$. For this, we apply an Empirical Bayes (EB) approach, where the estimate of $\eta$ is obtained by maximizing the marginal likelihood of the \emph{observed data} $w_{\bar\smY} = \begin{bmatrix}
	w_j^\top & w_a^\top
\end{bmatrix}^\top$, obtained by integrating out the dependence on the missing node data and the impulse response of the modules,
\begin{equation}\label{eq:2.1}
\hat\eta = \argmax_\eta p(w_{\bar\smY}; \eta).
\end{equation}

\begin{remark}
	If we do not consider additional node $w_a$, we can remove the extra layer of equation in $w_{\bar\smY}$. In the above model \eqref{eq:model}, it will be the second (block) row of equation in $w_{\bar\smY}$ and therefore we need not model $f$. Now the model will depend upon the vector of parameters,
\beqr
\eta := [ \
\theta \
{\lambda_j^s} \
\dots \
{\lambda_{jk_p}^s} \
{\lambda_{c_1}^b} \
\dots \
{\lambda_{mk_p}^b} \
&&{\beta_j^s} \
\dots \
{\beta_{jk_p}^s} \nonumber\\
&&{\beta_m^b} \
\dots \
{\beta_{mk_p}^b} \  \sigma_j^2 \
\sigma_m^2 ] . \nonumber
\eeqr
	We do not need to estimate extra parameters $\lambda_a^f$, ${\lambda_{c_1}^f}, \dots, {\lambda_{ak_p}^f}$, $\beta_a^f, {\beta_{c_1}^f}, \dots, {\beta_{ak_p}^f}, \sigma_{am}^2, \sigma_a^2$.
\end{remark}

The first important problem with the above approach of parameter $\eta$ inference in \eqref{eq:2.1} is that we need to deal with the unknown missing node observation $w_m$. Secondly, due to the incomplete model, the marginal pdf of $w_{\bar\smY}$ (i.e. $p(w_{\bar\smY};\eta)$) does not admit an analytical expression and cannot be computed under a closed-form solution. Adding to it, the maximization problem does not admit an explicit solution. In the next section, we study how to solve the marginal likelihood problem through a dedicated iterative scheme.

\section{Parameter Inference}\label{ML}
In this section, we provide the approach to deal with the missing node $w_m$ and the above discussed problems and solve the marginal likelihood problem in \eqref{eq:2.1}. We use the strategy of \emph{data augmentation} \cite{Tanner&Wong_ASA:1987, Vandyk&Meng_CGS:2001} to deal with the unknown missing node observations. In this data augmentation strategy, we treat the unknown node signal as auxiliary variables which are estimated along with the parameters in $\eta$. This data augmentation strategy has been used for state inference in identification of state-space models \cite{Schon&etal_SYSID:15}. There are various methods that use the data augmentation strategy like the EM algorithm \cite{Dempsteretal_JRSS:1977} for a Frequentist formulation of the identification problem and the \emph{Gibbs sampler} \cite{Gemen&Gemen_PA&MI:1984} for a Bayesian formulation.

For the problem in \eqref{eq:2.1}, which is a Frequentist formulation, we solve it by deriving an iterative solution scheme through the EM algorithm.
%In this section we focus on solving the problem in \eqref{eq:2.1}.
% Apart from the gradient based non-linear optimization solvers, algorithms for solving the hyper-parameter estimation problem	as in \eqref{eq:2.1} are investigated in \cite{chenalgorithms}.
%We use an iterative solution scheme through the EM algorithm.
For this, we need to first define the \emph{latent variables} whose estimation simplifies the computation of the marginal likelihood. The first natural choice is $w_m$, which is the missing node observation. Also, $s, b$ and $f$ are latent variables. Then, the solution to \eqref{eq:2.1} using the EM algorithm is obtained by iterating among the following two steps:
\begin{itemize}
	\item \emph{E-Step:} Given an estimate $\hat{\eta}^{(n)}$ computed at the $n^{th}$ iteration, compute
	\begin{equation}\label{eq:4.1}
	Q^{(n)}(\eta) = \mathbb{E}[\log p(w_{\bar\smY}, w_m, s, f, b ;\eta)] \,,
	\end{equation}
	% \begin{equation}\label{eq:4.1}
	% \begin{split}
	% \tilde Q^{(k)}(\eta) &= \mathbb{E}[\ln p(w_j, s_j, {s_{jk}}_1, {s_{jk}}_2, \dots, {s_{jk}}_p;\eta)]\\
	% \end{split}
	% \end{equation}
	where the expectation of the joint log-likelihood of $w_j , w_a, w_m, s, f$ and $b$ is taken with respect to the posterior $p(w_m, s, b, f| w_{\bar\smY};\hat{\eta}^{(n)})$;
	\item \emph{M-Step:} Update $\hat\eta$ by solving
	\begin{equation} \label{eq:Q_fun}
	\hat{\eta}^{(n+1)} = \argmax_\eta Q^{(n)}(\eta) \,.
	\end{equation}
\end{itemize}
When the two steps are iterated, convergence to a stationary point of the marginal likelihood (which can be a local minima or global minima) is ensured \cite{Boyles_JRSS:1983}. In the next section, we show the clear advantage of using the EM algorithm. We have transformed the original marginal likelihood problem \eqref{eq:2.1} to a sequence of problems that require solving \eqref{eq:Q_fun} using the EM algorithm. We show that, when we use the EM method, the nonlinear optimization problem becomes a problem of iteratively constructing analytical solutions and solving scalar optimization problems, which significantly simplifies solving \eqref{eq:2.1}.

Also, the E-step in the algorithm involves computing expectation with respect to the posterior distribution $p(w_m, s, p, f | w_{\bar\smY})$, which is
non-Gaussian and does not have an analytical form. Thus
the integral in \eqref{eq:4.1} is not tractable. In the next section, we present a solution to this problem by using a Markov Chain Monte Carlo (MCMC) method, \emph{Gibbs sampler}.

\subsection{Computation of E-step}
In order to perform the E-step we resort to the Monte Carlo approximation of \eqref{eq:4.1}. This method has been introduced in \cite{Wei&Tanner_ASA:1990} and is known as Monte Carlo Expectation Maximization (MCEM). In this, we approximate \eqref{eq:4.1} as,
\beq \label{eq:MCEM}
Q^{(n)}(\eta) \approx \frac{1}{M}\sum_{i = 1}^{M} \log p(w_{\bar\smY}, \bar w_m^{(i,n)}, \bar s^{(i,n)}, \bar p^{(i,n)}, \bar f^{(i,n)}; \eta) \, ,
\eeq
where $\bar s^{(i,n)}, \bar p^{(i,n)}, \bar f^{(i,n)}, \bar w_m^{(i,n)}$ are samples drawn at the $n^{th}$ iteration from the posterior $p(w_m, s, p,f|w_{\bar\smY};\hat{\eta}^{(n)})$. In order to draw samples from the posterior, we use the \emph{Gibbs sampler}\footnote{There are other joint posterior approximation techniques like Variational Bayes approximations \cite{Beal2003} and other MCMC methods \cite{Gilks&etal:1996}, which can also be applied. In this paper we resort to \emph{Gibbs sampler}. \emph{Gibbs sampler} does not require any tuning of proposal distribution and does not include any rejection step.}. The idea behind the \emph{Gibbs sampler}, which is a MCMC method, is to generate samples from a desired target distribution by simulating a Markov chain, with the target distribution as its stationary distribution. The Gibbs sampler produces samples from the posterior distribution by iteratively sampling each random variable conditioned on all other random variables \cite{Gemen&Gemen_PA&MI:1984}. Therefore to create samples from the joint posterior distribution, starting from an initialization $\bar s^{(0,n)}, \bar p^{(0,n)}, \bar f^{(0,n)}, \bar w_m^{(0,n)}$, we iteratively perform Algorithm \ref{algo:gibbs}
\begin{algorithm}[h]
	\begin{enumerate}
		\item sample $\bar w_m^{(i+1,n)} \sim p(w_m|w_j,\bar s^{(i,n)}, \bar b^{(i,n)},\bar f^{(i,n)})$ \ ,
		\item sample $\bar s^{(i+1,n)} \sim p(s|w_j, \bar w_m^{(i+1,n)}, \bar b^{(i,n)},\bar f^{(i,n)})$ \ ,
		\item sample $\bar b^{(i+1,n)} \sim p(b|w_j, \bar w_m^{(i+1,n)}, \bar s^{(i+1,n)},\bar f^{(i,n)})$ \ ,
		\item sample $\bar f^{(i+1,n)} \!\!\sim\!\! p(f|w_j, \bar w_m^{(i+1,n)}, \bar s^{(i+1,n)},\bar b^{(i+1,n)})$,
	\end{enumerate}
	\caption{Gibbs sampler}
	\label{algo:gibbs}
\end{algorithm}
%\begin{enumerate}\label{gibbs}
%	\item sample $\bar w_m^{(i+1,n)} \sim p(w_m|w_j,\bar s^{(i,n)}, \bar p^{(i,n)},\bar f^{(i,n)})$ \ ,
%	\item sample $\bar s^{(i+1,n)} \sim p(s|w_j, \bar w_m^{(i+1,n)}, \bar p^{(i,n)},\bar f^{(i,n)})$ \ ,
%	\item sample $\bar p^{(i+1,n)} \sim p(p|w_j, \bar w_m^{(i+1,n)}, \bar s^{(i+1,n)},\bar f^{(i,n)})$ \ ,
%	\item sample $\bar f^{(i+1,n)} \!\!\sim\!\! p(p|w_j, \bar w_m^{(i+1,n)}, \bar s^{(i+1,n)},\bar p^{(i+1,n)})$,
%\end{enumerate}
%\begin{enumerate}
%	\item sample $\bar w_m^{(i+1,n)} \sim p(w_m|w_j,\bar s^{(i,n)}, \bar p^{(i,n)})$ \ ,
%	\item sample $\bar s^{(i+1,n)} \sim p(s|w_j, \bar w_m^{(i+1,n)}, \bar p^{(i,n)})$ \ ,
%	\item sample $\bar p^{(i+1,n)} \sim p(p|w_j, \bar w_m^{(i+1,n)}, \bar s^{(i+1,n)})$ \ ,
%\end{enumerate}
for a large number of iterations keeping the hyperparameters value fixed. Normally, we discard first few samples since the Markov chain will be poorly mixed and the obtained samples will be far away from the stationary distribution, which is the target distribution for the \emph{Gibbs sampler}. Therefore, we discard the first $B$ samples, and this is known as \emph{burn-in} period. If the burn-in period is large enough, then we produce samples that come from the stationary distribution\footnote{The choice of burn-in period is a non-trivial problem which is outside the scope of this paper and methods to address this problem have been provided in \cite{Gilks&etal:1996}.}.
Similar to the other MCMC techniques, the generated samples can be correlated, but we need independent samples. Instead of sampling individual variables, a group of variables can be sampled to tackle this. This is called \emph{blocking Gibbs sampling} algorithm or \emph{blocked Gibbs sampler} \cite{Jensen&etal_IJHCS:1995}. This is done by choosing blocks
of variables, not necessarily disjoint, and then sampling jointly from the variables in
each block in turn, conditioned on the remaining variables. We adopt this approach of Gibbs sampling in this paper.
Another approach called \emph{thinning} can be used to reduce the correlation in generated samples, where after the burn-in period each sample can be collected after $\kappa$ iterations.

It is important to note that in order to use the Gibbs sampler, the above conditional distributions should be known and we should be able to generate samples from them. This is a requirement when using Gibbs sampling. Gibbs sampling is a simple and effective sampling method, provided we know the conditional distributions. Next we show that, in our situation, the conditional distributions that we require have a convenient form.

\begin{proposition}\label{prop2}
	Consider the model in \eqref{eq:model}. The conditional distributions of $s, p, f, w_m$ are Gaussian and given by,
	\begin{align}
	p(w_m|w_{\bar\smY},s, b, f) & \sim \mathcal{N}(\mu_w, P_w) \, , \label{eq:posterior_wm} \\
	p(s|w_{\bar\smY},w_m, b, f) & \sim \mathcal{N}(\mu_s, P_s) \, , \label{eq:posterior_s} \\
	p(b|w_{\bar\smY},w_m, s, f) & \sim \mathcal{N}(\mu_b, P_b) \, , \label{eq:posterior_p} \\
	p(f|w_{\bar\smY},w_m, s, b) & \sim \mathcal{N}(\mu_f, P_f) \, , \label{eq:posterior_f}
	\end{align}
	where,
	\begin{align}
	P_w &= ({\bar\mu_2^\top\Sigma^{-1}\bar\mu_2} + \Lambda_{22} - \begin{bmatrix}
		\Lambda_{21} & \Lambda_{22}
	\end{bmatrix}\bar\mu_2 \nonumber\\
	&\quad\quad\quad\quad\quad\quad\quad\quad\quad\quad- \bar\mu_2^\top\begin{bmatrix}
	\Lambda_{12}^\top & \Lambda_{22}^\top
	\end{bmatrix}^\top )^{-1} \, ,\\
	\mu_w &= P_w(\bar\mu_2^\top \begin{bmatrix}
		\Lambda_{11} & \Lambda_{12}
	\end{bmatrix}^\top w_{\bar\smY} + \begin{bmatrix}
	\Lambda_{21} & \Lambda_{22}
	\end{bmatrix} \bar\mu_1  \nonumber\\
	&\quad\quad\quad\quad\quad\quad\quad\quad\quad-\Lambda_{21}w_{\bar\smY} - \bar \mu_2^\top \Sigma^{-1}\bar\mu_1),\\
	P_s &= \left(\mathbf K_1^{-1} + \bar{\mathbf W}^\top{\Sigma^{-1}}\bar{\mathbf W}\right)^{-1}, \\
	\mu_s &= P_s\bar{\mathbf W}^\top{\Sigma^{-1}}\left(w_{\smY} - \bar \mu_4\right),\\
	P_b &= \left(\mathbf K_2^{-1} + \bar{\mathbf R}^\top{\Sigma^{-1}}\bar{\mathbf R}\right)^{-1},
	\end{align}
	\begin{align}
	\mu_b &= P_b\bar{\mathbf R}^\top{\Sigma^{-1}}\left(w_{\smY} - \bar \mu_5\right),\\
	P_f &= \left(\mathbf K_3^{-1} + \bar{\mathbf Q}^\top{\Sigma^{-1}}\bar{\mathbf Q}\right)^{-1}, \\
	\mu_f &= P_f\bar{\mathbf Q}^\top{\Sigma^{-1}}\left(w_{\smY} - \bar \mu_6\right).
	\end{align}
\end{proposition}
\vspace{-0,5cm}
{\bf Proof:} Collected in the appendix. The expressions for $\bar{\mathbf W}, \bar{\mathbf R}, \bar{\mathbf Q}, \bar \mu_{1}, \bar \mu_2, \bar \mu_4, \bar \mu_5$ and  $\bar \mu_6$ are provided in the appendix.
%\begin{proposition}\label{prop2}
%	Consider the model in \eqref{eq:model}. The conditional distributions of $s, p, w_m$ are Gaussian and given by,
%	\begin{align}
%	p(w_m|w_j,s, p) & \sim \mathcal{N}(\mu_w, P_w) \, , \label{eq:posterior_wm} \\
%	p(s|w_j,w_m, p) & \sim \mathcal{N}(\mu_s, P_s) \, , \label{eq:posterior_s} \\
%	p(p|w_j,w_m, s) & \sim \mathcal{N}(\mu_p, P_p) \, , \label{eq:posterior_p}
%	\end{align}
%	where
%	\begin{align}
%P_s &= \left(\mathbf K_\beta^{-1} + \frac{\mathbf W^\top\mathbf W}{\sigma_j^2} \right)^{-1}\, , \\
%\mu_s &= \frac{P_s\mathbf W^\top}{\sigma_j^2}\left(w_j - \mu_4\right)\, ,\\
%%\mu_4  &= W_{ji}g_{ji} + \ell\left(w_m - \mathbf Rm - r_m \right) + r_j \\
%	P_w &= \left(\frac{1}{\sigma_m^2}\bC + \frac{\mu_2^\top\mu_2}{\sigma_j^2}\right)^{-1}\, ,\\
%%	C &= (I - \bar P_m)^\top(I - \bar P_m),\\
%	\mu_w &= \bC^{-1}(\mu_3-\bar P_m^\top\mu_3)+\frac{P_w\mu_2^\top}{\sigma_j^2}(w_j - \mu_{11})\, ,\\
%%	\mu_{11} &= \mu_1 + \mu_2C^{-1}(\mu_3-\bar P_m^\top\mu_3)\\
%	P_p &= \left(\mathbf K_\gamma^{-1} + \ell^2\frac{\mathbf R^\top\mathbf R}{\sigma_j^2} \right)^{-1}\, ,\\
%	\mu_p &= -\frac{\ell P_p\mathbf R^\top}{\sigma_j^2}\left(w_j - \mu_5\right).
%	\end{align}
%	{\bf Proof:} Collected in the appendix. The closed form expression for $\bC, \mu_{11}, \mu_2, \mu_3, \mu_4$ and  $\mu_5$ are provided in the appendix. \hfill \qed
%\end{proposition}

Therefore, it is easy to set up the Gibbs sampler and sample from the joint posterior distribution, thereby approximating \eqref{eq:4.1} using \eqref{eq:MCEM}.
\begin{remark}
	When we do not consider additional node $w_a$, we do not consider the extra layer of equation in $w_{\bar\smY}$ (i.e. $w_{\smY}$ becomes $(w_j,w_m)$) and we need not model $f$. Therefore, we discard the use of $f$ and expressions related to it.
%	\begin{enumerate}
%		\item sample $\bar w_m^{(i+1,n)} \sim p(w_m|w_j,\bar s^{(i,n)}, \bar b^{(i,n)})$ \ ,
%		\item sample $\bar s^{(i+1,n)} \sim p(s|w_j, \bar w_m^{(i+1,n)}, \bar b^{(i,n)})$ \ ,
%		\item sample $\bar b^{(i+1,n)} \sim p(b|w_j, \bar w_m^{(i+1,n)}, \bar s^{(i+1,n)})$ \ .
%	\end{enumerate}
	The conditional distributions follow the same equations as above.
\end{remark}

\subsection{Computation of M-step}\label{mstep}
Next we move to the M-step where we update the vector of parameters according to \eqref{eq:Q_fun}. We need to maximize \eqref{eq:MCEM} with respect to the vector of parameters in $\eta$. We will now show that the optimization problem can be split into several independent optimization problems that depend on different components of the vector of parameters $\eta$. We can split the optimization problem as,
\beqr
&&Q^{(n)}(\eta) \nonumber\\
&&\quad= \argmax_\eta\frac{1}{M}\!\!\sum_{i = 1}^{M} \log p(w_{\bar\smY}, \bar s^{(i,n)}, \bar b^{(i,n)}, \bar f^{(i,n)}, \bar w_m^{(i,n)}; \eta) \nonumber\\
&&\quad= A + B + C + D \, ,\label{eq:optisplit}
\eeqr
where,
\beqr
A &=&\!\argmax_{\theta,\bar\Sigma}\!\!\frac{1}{M}\!\!\sum_{i = 1}^{M} \log p(w_{\bar\smY}, \bar w_m^{(i,n)}|\bar s^{(i,n)}, \bar b^{(i,n)}, \bar f^{(i,n)}) \nonumber\\
B &=& \argmax\frac{1}{M}\sum_{i = 1}^{M}\bigg[\sum_{k \in \mathcal{D}_m^w} \log p(\bar b_{k}^{(i,n)}; \lambda_{k}^b,\beta_{k}^b) \nonumber\\
&&+ \log p(\bar b_{m}^{(i,n)}; \beta_{m}^b) + \sum_{k \in \mD_m^u} \log p(\bar b_{mk}^{(i,n)}; \lambda_{mk}^b,\beta_{mk}^b)\bigg]\nonumber\\
C &=& \argmax\frac{1}{M}\sum_{i = 1}^{M}\bigg[\sum_{k \in j \cup \mathcal{D}_j^w \backslash \{i,m\}} \!\!\!\!\log p(\bar s_{k}^{(i,n)}; \lambda_{k}^s,\beta_{k}^s) \nonumber\\
&&+ \log p(\bar s_{m}^{(i,n)}; \beta_{m}^s) + \sum_{k \in \mathcal{D}_j^u \backslash \{j\}} \!\!\!\!\log p(\bar s_{jk}^{(i,n)}; \lambda_{jk}^s,\beta_{jk}^s)\bigg]\nonumber\\
D &=& \argmax\frac{1}{M}\sum_{i = 1}^{M}\bigg[\sum_{k \in a \cup \mathcal{D}_a^w\backslash\{m\}} \log p(\bar f_{k}^{(i,n)}; \lambda_{k}^f,\beta_{k}^f) \nonumber\\
&&+ \log p(\bar f_{m}^{(i,n)};\beta_{m}^f)+ \sum_{k \in \mathcal{D}_a^u} \log p(\bar f_{ak}^{(i,n)}; \lambda_{ak}^f,\beta_{ak}^f)\bigg]\nonumber \, ,
%\label{eq:optisplit}
\eeqr
%\beqr
%Q^{(n)}(\eta) &=& \argmax_\eta\frac{1}{M}\sum_{i = 1}^{M} \log p(w_j, \bar s^{(i,n)}, \bar p^{(i,n)}, \bar w_m^{(i,n)}; \eta) \nonumber\\
%&=& \argmax_{\ell,\theta,\sigma_j^2}\frac{1}{M}\sum_{i = 1}^{M} \log p(w_j|\bar s^{(i,n)}, \bar p^{(i,n)}, \bar w_m^{(i,n)}) \nonumber\\
%&+& \argmax_{\sigma_m^2}\frac{1}{M}\sum_{i = 1}^{M} \log p(\bar w_m^{(i,n)}|\bar p^{(i,n)};\sigma_m^2)\nonumber\\
%&+& \argmax_{\gamma}\frac{1}{M}\sum_{i = 1}^{M} \log p(\bar p^{(i,n)}; \gamma) \nonumber\\
%&+& \argmax_{\beta}\frac{1}{M}\sum_{i = 1}^{M} \log p(\bar s^{(i,n)}; \beta) \nonumber\\
%&=& A + B + C + D \, ,\label{eq:optisplit}
%\eeqr
and $\bar\Sigma$ represent the parameters in the covariance matrix $\sigma_j^2, \sigma_m^2, \sigma_a^2, \sigma_{am}^2$.

\subsubsection{Update of kernel hyperparameters}
From \eqref{eq:optisplit}, we can see that the hyperparameters of each kernel can be updated independently from the rest of the parameters in $\eta$. The following proposition provides a means to update the kernel hyperparameters, except the hyperparameters of the kernel for which $\lambda$'s are set to 1.

\begin{proposition}\label{prop3ab}
Let $$\hat s_k^{(n)} = \frac{1}{M}\sum_{i = 1}^{M} \bar s_k^{(i,n)},$$ $$\hat{\mathbf S}_k^{(n)} = \frac{1}{M}\sum_{i = 1}^{M} (\bar s_k^{(i,n)}-\hat s_k^{(n)})(\bar s_k^{(i,n)}-\hat s_k^{(n)})^\top,$$ and analogously define $\hat b_k^{(n)}, \hat f_k^{(n)}, \hat{\mathbf B}_k^{(n)}, \hat{\mathbf F}_k^{(n)}$. Define
\beqr\label{eq:101}
\!\!\!{Q_\beta}_k^{(n)}&&({\beta_k}) = \log \det({{K_\beta}_k}) \nonumber \\
&&\!+ l \log \!\bigg(\hat s_k^{\top(n)}{({{K_\beta}_k})}^{-1}\hat s_k^{(n)} \!+\! \mathrm{tr}\big({({{K_\beta}_k})}^{-1}\hat{\mathbf S}_k^{(n)}\big)\!\bigg)
\eeqr
for $k = \{j, c_1, \dots, jk_p\}$ where $c_1,\,\ldots,\,c_p$ and $k_1,\,\ldots,\,k_p$ are the elements of the set $\mathcal{D}_j^w \backslash \{i \}$ and $\mD_j^{u}\backslash\{j\}$ respectively.
% $k \in \{\mathcal N_j\cup j\} \backslash i$.
%, and
% \begin{equation}
% {Q_\beta}_j^{(n)}({\beta_j}) = \log \det({{K_\beta}_j}) + l \log \bigg\{\mathrm{tr}\big({({{K_\beta}_j})}^{-1}\hat{\mathbf S}_j^{(n)}\big)\bigg\}.
% \end{equation}
Then,
\begin{equation}\label{eq:12.1}
{\hat{\beta_k}}^{(n+1)} = \argmin_{{\beta_k} \in [0,1)}{Q_\beta}_k^{(n)}({\beta_k}); \end{equation}
% \begin{equation}\label{eq:12.2}
% {\hat{\beta_j}}^{(n+1)} = \argmin_{\beta_j \in [0,1)}{Q_\beta}_j^{(n)}({\beta_j})
% \end{equation}
\begin{equation}\label{eq:13.1}
{\hat{\lambda_k}}^{(n+1)} \!=\! \frac{1}{l}(\hat s_k^{\top(n)}{(K_{{\hat\beta}_{{k}}^{(n+1)}})}^{-1}\hat s_k^{(n)} \!+\! \mathrm{tr}\big({(K_{{\hat\beta}_{{k}}^{(n+1)}})}^{-1}\hat{\mathbf S}_k^{(n)}\big)).
\end{equation}
The updates for $\beta$ and $\lambda$ for impulse responses $b_k, k \in m \cup \mathcal{D}_m^w$, $b_{mk}, k \in \mD_m^u$, and $f_k, k \in a \cup \mathcal{D}_a^w$, $f_{ak}, k \in \mD_a^u$  are updated analogously by using $\hat b_k^{(n)}, \hat{\mathbf B}_k^{(n)}$ and $\hat f_k^{(n)}, \hat{\mathbf F}_k^{(n)}$ respectively.
\end{proposition}
{\bf Proof:} See the appendix.	

To tackle the identifiability issues, we have fixed the hyperparameters $\lambda_m^s$, $\lambda_{m}^b$ and $\lambda_{m}^f$ (i.e. $\lambda$'s corresponding to the modules having missing node as inputs) to be 1. We now provide means to update the respective $\beta$ hyperparameters of the kernel of the corresponding modules using the following proposition.
\begin{proposition}\label{prop3b}
	The updates of kernel's hyperparameters related to the impulse response of modules with missing node as inputs are obtained by solving the scalar optimization problem in the domain $[0,1)$,
	\beqr\label{eq:beta1}
	\hat\beta_{m}^{s(n+1)} &=& \argmin_{\beta_{m}^s} \log \det({K_{\beta_m^s}}) \nonumber\\
	&& + \hat s_{m}^{\top(n)}{({K_{\beta_m^s}})}^{-1}\hat s_{m}^{(n)} \!+\! \mathrm{tr}\big({({K_{\beta_m^s}})}^{-1}\hat{\mathbf S}_{m}^{(n)}\big) \, .
	\eeqr
	The updates for $\beta_m^b$ and $\beta_{m}^f$ for impulse responses $b_m$ and $f_{m}$ are updated analogously by using $\hat b_m^{(n)}, \hat{\mathbf B}_m^{(n)}$ and $\hat f_m^{(n)}, \hat{\mathbf F}_m^{(n)}$ respectively.
\end{proposition}
{\bf Proof:} See the appendix.	

% \begin{proposition}\label{prop3}
% 	The updates of kernel's hyperparameters are obtained by solving the scalar optimization problem in the domain $[0,1)$,
% 	\beqr\label{eq:beta}
% 	\hat\beta_k^{(n+1)} =\argmin_{\beta_k} \frac{1}{M}&&\sum_{i = 1}^{M} \bigg[ \log[\det({{K_\beta}_k})] \nonumber\\
% 	&+& {\bar s_{k}^{(i,n)\top}}{({{K_\beta}_k})}^{-1}{\bar s_{k}^{(i,n)}} \bigg] \, ,
% 	\eeqr	
% 	for  $k \in \{\mathcal N_j \cup j\} \backslash i$ and
% 	\beqr\label{eq:gamma}
% 	\hat\gamma_k^{(n+1)} = \argmin_{\gamma_k} \frac{1}{M}&&\sum_{i = 1}^{M} \bigg[ \log[\det({{K_\gamma}_k})] \nonumber\\
% 	&+& {\bar p_{k}^{(i,n)\top}}{({{K_\gamma}_k})}^{-1}{\bar p_{k}^{(i,n)}} \bigg] \, ,
% 	\eeqr
% 	for $k \in \{\mathcal{D}_m \cup \mU_m^r\}$ and
% 	\beqr\label{eq:delta}
% 	\hat\delta_k^{(n+1)} = \argmin_{\delta_k} \frac{1}{M}&&\sum_{i = 1}^{M} \bigg[ \log[\det({{K_\delta}_k})] \nonumber\\
% 	&+& {\bar f_{k}^{(i,n)\top}}{({{K_\delta}_k})}^{-1}{\bar f_{k}^{(i,n)}} \bigg] \, ,
% 	\eeqr
% 	for $k \in \{\mathcal{D}_a \cup \mU_a^r\}$.
% \end{proposition}
% {\bf Proof}: Expanding the pdf in the optimization problem $C$ (and $D, E$) in \eqref{eq:optisplit}, we get the above expressions. \hfill \qed

\begin{remark}
	The optimization problem in \eqref{eq:12.1} and \eqref{eq:beta1} can be difficult to perform in practice when the determinant of the kernel has a very low value or when the inversion of the kernel becomes difficult. To tackle this, we exploit the factorization of the \emph{first order stable spline kernel} as in \cite{Bottegaletal_Autom:16} by writing ${K_\beta}_k = L D(\beta) L^T$, where $L$ is lower-triangular with known entries (essentially, an ``integrator'') and $D(\beta)$ is diagonal with entries essentially being an exponential functions of $\beta$. Using the above technique also increases the computation speed of the algorithm.
\end{remark}
\begin{remark}
	When we do not consider an additional node $w_a$, we need to update only the hyperparameter $\lambda$'s and $\beta$'s related to impulse responses in $s,b$.
\end{remark}
%\begin{proposition}\label{prop3}
%	The updates of kernel's hyperparameters are obtained by solving the scalar optimization problem in the domain $[0,1)$,
%	\beqr\label{eq:beta}
%	\hat\beta_k^{(n+1)} =\argmin_{\beta_k} \frac{1}{M}&&\sum_{i = 1}^{M} \bigg[ \log[\det({{K_\beta}_k})] \nonumber\\
%	&+& {\bar s_{k}^{(i,n)\top}}{({{K_\beta}_k})}^{-1}{\bar s_{k}^{(i,n)}} \bigg] \, ,
%	\eeqr	
%	for  $k \in \{\mathcal N_j \cup j\} \backslash i$ and
%	\beqr\label{eq:gamma}
%	\hat\gamma_k^{(n+1)} = \argmin_{\gamma_k} \frac{1}{M}&&\sum_{i = 1}^{M} \bigg[ \log[\det({{K_\gamma}_k})] \nonumber\\
%	&+& {\bar p_{k}^{(i,n)\top}}{({{K_\gamma}_k})}^{-1}{\bar p_{k}^{(i,n)}} \bigg] \, ,
%	\eeqr
%	for $k \in \{\mathcal{D}_m \cup \mU_m^r\}$.
%\end{proposition}
%{\bf Proof}: Expanding the pdf in the optimization problem $C$ (and $D$) in \eqref{eq:optisplit}, we get the above expressions. \hfill \qed

\subsubsection{Update of $\theta$ and noise covariance}
Following \eqref{eq:optisplit}, the updates of $\theta$ and the noise covariance parameters in $\eta$ are independent of the kernel hyperparameters. Following a reasoning similar to \cite{Astrom_Autom:1980}, $\theta$ and $\Sigma$ are updated as per the \textit{Proposition \ref{prop5}}.
\begin{proposition} \label{prop5}
	Let
% 	\footnote{dependence on $\theta$ is dropped for convenience.}
	$\bar\varepsilon_j^{(i,n)}(\theta) = w_{j} - W_{ji}g_{ji}(\theta) - u_j - \bar{\mathbf W}^{(i,n)}\bar s^{(i,n)}$, $\tilde\Sigma =  \begin{bmatrix}
		\sigma_{a}^2 & \sigma_{am}^2 \\
		\sigma_{am}^2 & \sigma_{m}^2
	\end{bmatrix}$, and
	\beqr
	\bar\varepsilon^{(i,n)}(t) &&= \begin{bmatrix}
		w_a(t) \\ \bar w_m^{(i,n)}(t)
	\end{bmatrix} - \begin{bmatrix}
u_a(t) \\ u_m(t)
\end{bmatrix}  \nonumber\\
&&\quad\quad- \begin{bmatrix}
	\bar{\mathbf Q}^{(i,n)}(t, \star) & \mathbf 0 \\
	\mathbf 0 & \bar{\mathbf R}^{(i,n)}(t, \star) \\
\end{bmatrix}\begin{bmatrix}
	\bar f^{(i,n)} \\ \bar b^{(i,n)}
\end{bmatrix} \, , \nonumber
\eeqr
where $\bar{\mathbf Q}^{(i,n)}(t, \star)$, $\bar{\mathbf R}^{(i,n)}(t, \star)$ corresponds to the $t^{th}$ row of the matrix ${\mathbf Q}$, ${\mathbf R}$ respectively, with $w_m$ in the matrices substituted with $\bar w_m^{(i,n)}$. $\bar{\mathbf W}^{(i,n)}$ corresponds to the matrix ${\mathbf W}$, with $w_m$ in the matrices substituted with $\bar w_m^{(i,n)}$. Define $$\hat \varepsilon^{(n)}(t) = \frac{1}{M}\sum_{i = 1}^{M} \bar\varepsilon^{(i,n)}(t),$$ $$\hat{\mathbf E}^{(n)}(t) = \frac{1}{M}\sum_{i = 1}^{M} (\bar\varepsilon^{(i,n)}(t) -\hat \varepsilon^{(n)}(t))(\bar\varepsilon^{(i,n)}(t) -\hat \varepsilon^{(n)}(t))^\top.$$
% \beqr
% 	Q_{\theta}^{(n)}(\theta) &=& \norm{\bar\varepsilon_j^{(i,n)}(\theta)} \ \nonumber.
% 	\eeqr
	Then
	\begin{equation}\begin{aligned}
    \hat{\theta}^{(n+1)} & = \argmin_\theta \bigg[g_{ji}^\top \hat A^{(n)}g_{ji} - 2\hat{b}^{(n)\top}g_{ji}\bigg] \,,\\
    % \end{aligned}\end{equation}
    % \begin{equation}\begin{aligned}
% 	\hat{\theta}^{(n+1)} & = \argmin_{\theta} \frac{1}{M}\sum_{i = 1}^{M}\norm{\bar\varepsilon_j^{(i,n)}(\theta)}^2\ ,\\
	\hat \sigma_j^{2(n+1)} & = \frac{1}{NM}\sum_{i = 1}^{M}\norm{\bar\varepsilon_j^{(i,n)}(\hat\theta^{n+1})}^2,\\
	\hat{\tilde\Sigma}^{(n+1)} & = \frac{1}{N}(\sum_{t=1}^{N} [\hat \varepsilon^{(n)}(t,\hat\theta^{(n+1)})\hat \varepsilon^{(n)}(t,\hat\theta^{(n+1)})^\top \!\!+\!\! \hat{\mathbf E}^{(n)}(t)]) \ .
	\end{aligned}\label{eq:thesigopt}\end{equation}
\end{proposition}
{\bf Proof:} See the appendix.	

\begin{remark}
If $g_{ji}$ is linearly parameterized in $\theta$ (e.g. in case of FIR models), the above problem related to the update of $\theta$ becomes quadratic and a closed-form solution is achieved. That is, if $g_{ji} = M\theta$ where $M \in \mathbb{R}^{N\times n_{\theta}}$, then
\begin{equation}\label{eq:18}
\hat{\theta}^{(n+1)} = \big(M^\top \hat A^{(n)}M\big)^{-1}M^\top\hat{b}^{(n)}.
\end{equation}
\end{remark}
% The sparse elements in $\Sigma$ that are known apriori can be updated in $\hat{\Sigma}^{(n+1)}$. \giulio{[This is not clear: Better to say that the update of Sigma is that matrix on the right hand side of the equation, Hadarmard-multiplied by
% $$
% \begin{bmatrix}
%     1 & 0 & 0 \\ 0 & 1 & 1 \\ 0 & 1 & 1
% \end{bmatrix}
% $$]}
Note that, as shown in \cite{Ramaswamy&etal_Autom:2021}, we can update the parameter of the target module (i.e. $\theta$) using similar analytical expression as in \eqref{eq:18} for any other rational model structures as well (e.g. BJ models). This can be done by following the similar approach of \cite{Ramaswamy&etal_Autom:2021}.
From \eqref{eq:thesigopt}, we can observe that the update of $\bar\Sigma$ in each iteration of the MCEM algorithm is a closed form analytical solution and the update of $\theta$ is a nonlinear least-squares problem with decision variables being the parameters of the target module which are fewer than a direct PEM that includes the nuisance modules parameters as well in the problem as decision variables. Also the result of Propositions \ref{prop3ab} and \ref{prop3b} show that the update of kernel hyperparameter $\beta$'s are scalar optimization problems and $\lambda$'s have closed form solutions. Therefore, we have obtained a fast iterative procedure that follows simple rules to update the parameters, and provides a local solution to the marginal likelihood problem \eqref{eq:2.1} under the presence of missing node observations. Algorithm \ref{algo:complete} summarizes the steps to follow to obtain $\hat \eta$ and therefore $\hat \theta$.
\begin{algorithm}[h]\label{algo:3}
	% \SetAlgoLined
%	\textbf{Input:} $\{w_k(t)\}_{t=1}^N$, $k =1,\ldots,p$ \\
%	\textbf{Output:} $\hat \theta$
	% \Do{$\frac{\norm{\hat{\eta}^{(n)} - \hat{\eta}^{(k-1)}}}{\norm{\hat{\eta}^{(k-1)}}} > \mathrm{threshold}$}{
	\begin{enumerate}
		\item Set $n = 0$, Initialize $\hat{\eta}^{(0)}$.
		\item Run Gibbs sampler according to Algorithm \ref{algo:gibbs} and collect $M$ samples after discarding first $B$ samples for burn-in period using the result of Proposition \ref{prop2}.
		\item Update kernel hyperparameters using the result of Proposition \ref{prop3ab} and \ref{prop3b}.
		\item Update $\hat{\theta}^{(n+1)}$ and $\hat{\Sigma}^{(n+1)}$ using result of proposition \ref{prop5}.
		\item Update $\hat{\eta}^{(n+1)}$ using the above updated values of the parameters.
%		\item Set $\hat{\eta}^{(n+1)}\\ = [\begin{smallmatrix}
%		\hat{\theta}^{\top(n+1)} &
%		{\hat{\lambda_j}}^{(n+1)}  &
%		{\hat{\lambda_k}}_1^{(n+1)}  &
%		\dots &
%		{\hat{\lambda_k}}_p^{(n+1)}
%		\end{smallmatrix}\\
%		\begin{smallmatrix}
%		\quad \quad \quad \quad \quad \quad \quad \quad {\hat{\beta_j}}^{(n+1)}  &
%		{\hat{\beta_k}}_1^{(n+1)}  &
%		\dots &
%		{\hat{\beta_k}}_p^{(n+1)}  &
%		{\hat{\bar\sigma}}_j^{2(n+1)}
%		\end{smallmatrix}]^\top$
		\item Set $n = n + 1$.
		\item Repeat from steps (2) to (8) until convergence.
	\end{enumerate}
	%Extract the estimated model parameters $\hat\theta$ from the converged $\hat\eta$ to get the dynamic model $G_{ji}$
	\caption{Algorithm for local module identification in dynamic networks under missing node observations}
	\label{algo:complete}
\end{algorithm}

%\vspace{-0.6cm}
The initialization can be done by randomly choosing $\eta$ considering the constraints of the hyperparameters. The convergence criterion for the algorithm depend on the value of $\frac{\norm{\hat{\eta}^{(n)} - \hat{\eta}^{(n-1)}}}{\norm{\hat{\eta}^{(n-1)}}}$. This value should be small for convergence so that the algorithm can be terminated. A value of $10^{-2}$ is considered for the numerical simulations in Section \ref{sec:Num_ex}. The other convergence criterion is the maximum number of iterations. It is taken as 50. For the numerical simulations, the initialization of the latent variables ($s, p, f, w_m$) for the Gibbs sampler are taken as a zero vector. The number of samples $M$ for the Gibbs sampler is taken as 100 and the burn-in period $B$ is equal to 2000.

\begin{remark}
	The above developed method can be applied when we have the input of the target module (i.e. $w_i$) as the missing node observation with slight modifications in the above results. However, in this case, we might face identifiability issues of the target module. This is because of the fact that we can estimate the missing node signal and the target module up to a scaling factor. This has been a common issue in blind system identification \cite{Abedetal_IEEE:1997}.
\end{remark}

\section{Numerical simulations}\label{sec:Num_ex}
Numerical simulations are performed to evaluate the performance of the developed method. The simulations are performed on the dynamic network depicted in Figure \ref{fig:dynnet_Ex_wnoise1_10}. The goal is to identify $G_{31}^0$, which is the target module. The modules of the network in Figure \ref{fig:dynnet_Ex_wnoise1_10} are given by,
% \giulio{replace $z$ with $q$}
\begin{align*}
	&G_{31}^0 = \frac{q^{-1} + 0.05q^{-2}}{1 + q^{-1} + 0.6q^{-2}} = \frac{b_1^0q^{-1} + b_2^0q^{-2}}{1 + a_1^0q^{-1} + a_2^0q^{-2}}\\
	&G_{32}^0 = \frac{0.225 q^{-1}}{1 + 0.5 q^{-1}};\\
	&G_{34}^0 = \frac{1.184 q^{-1} - 0.647 q^{-2} + 0.151 q^{-3} - 0.082 q^{-4}}{1 - 0.8 q^{-1} + 0.279 q^{-2} - 0.048 q^{-3} + 0.01 q^{-4}};\\
	&G_{14}^0 = G_{21}^0 = \frac{0.4q^{-1} - 0.5q^{-2}}{1 + 0.3q^{-1}};H_{1}^0 = \frac{1}{1 + 0.2q^{-1}};\\
	&G_{12}^0 = G_{24}^0 = \frac{0.4q^{-1} + 0.5q^{-2}}{1 + 0.3q^{-1}}; H_{2}^0 = \frac{1}{1 + 0.3q^{-1}};\\
	&H_{4}^0 = 1; H_{3}^0 = \frac{1 - 0.505 q^{-1} + 0.155 q^{-2} - 0.01 q^{-3}}{1 - 0.729 q^{-1} + 0.236 q^{-2} - 0.019 q^{-3} }.
\end{align*}
For estimation $G_{31}^0$ using the direct method, we need to measure $w_1$, $w_2$, $w_3$ and $w_4$ and solve a 3-input/1-output MISO identification problem with $w_1(t)$, $w_2(t)$ and $w_4(t)$ as inputs. $w_2$ needs to be included as predictor input in order to satisfy the parallel path/ loop condition (i.e. Condition \ref{condx1}) and $w_4$ needs to be included as predictor input to satisfy the confounding variable condition. Now, we consider the case where we cannot measure $w_2$, which leads to lack of consistency in the direct method since we cannot satisfy the parallel path/loop condition. In this case, we resort to the approach developed in this paper and resort to the following options:
\begin{enumerate}
	\item consider $w_2$ as missing node, i.e. $w_m = w_2$, and consider the predictor model with $w_{\smY} = \{w_3, w_2\}$. We add to $w_{\smD}$ the measured node signals that have unmeasured paths to $w_{\smY}$ (i.e. $w_1, w_4$) and the missing node signal $w_2$. Therefore $w_{\smD} = \{w_1, w_4, w_2\}$, $w_{\smW} = \{w_1, w_2, w_3, w_4\}$, $w_{\smO} = \{w_3\}$ and $w_{\smQ} = \{w_2\}$. By this signal selection , we can verify that the conditions \ref{condx2ab}, \ref{cond:path_OtoJ} and \ref{cond5} are satisfied. This identification strategy is mentioned below as MC-EBDM;
	\item consider $w_2$ as missing node and add the descendant $w_1$ of $w_2$ as additional output, i.e. $w_m = w_2$, $w_a = w_1$ and consider the model with $w_{\smY} = \{w_3, w_2, w_1\}$. We add to $w_{\smD}$ the measured node signals that have unmeasured paths to $w_{\smY}$ (i.e. $w_1, w_4$) and the missing node signal $w_2$. Therefore $w_{\smD} = \{w_1, w_4, w_2\}$, $w_{\smW} = \{w_1, w_2, w_3, w_4\}$, $w_{\smO} = \{w_3\}$ and $w_{\smQ} = \{w_1, w_2\}$. By this signal selection , we can verify that the conditions \ref{condx2ab}, \ref{cond:path_OtoJ} and \ref{cond5} are satisfied. This identification strategy is mentioned below as MC-EBDMA.
\end{enumerate}

We compare the following identification strategies:
 \begin{description}
 \item[MC-EBDM] This is the method developed in this paper, namely Empirical Bayes Direct method with Monte Carlo sampling to deal with missing nodes; in particular, this estimator does not use additional node(s) and considers the predictor model $\{w_1, w_2, w_4\} \rightarrow \{w_3, w_2\}$;
 \item[MC-EBDMA] This is a variant of MC-EBDM that uses $w_1$ as additional node; and considers the predictor model $\{w_1, w_2, w_4\} \rightarrow \{w_1, w_2, w_3\}$
 \item[EBDM+M] This is the EBDM method developed in \cite{Ramaswamy&etal_Autom:2021}; this estimator does not encompass missing nodes and considers the predictor model $\{w_1, w_4\} \rightarrow \{w_3\}$;in other words it discards the non-measured node signal $w_2$;
 \item[EBDM] This is the same estimator as the previous one, with the assumption that the missing node $w_2$ is measurable (oracle assumption). We use this estimator as an upper bound of the performance of our developed method to reconstruct the missing node observation and identify the target module. Therefore, it considers a predictor model $\{w_1, w_2, w_4\} \rightarrow \{w_3\}$ where $w_2$ is known;
 \item[DM+TO] This is the standard MISO direct method first proposed in \cite{VandenHof&etal_Autom:13}, with the assumption that the missing node $w_2$ is measurable (oracle assumption). Therefore, it considers a predictor model $\{w_1, w_2, w_4\} \rightarrow \{w_3\}$ where $w_2$ is known and assumes a fully parametric model structure. Note that in order to avoid biased target module estimates in the direct method framework, we need $w_2$ to be measured and included as one of the predictor inputs \cite{Dankers&etal_TAC:16};
 \item[DM+TO+M] This is the same estimator as the previous one; it assumes a fully parametric model structure and has no specific way to deal with missing nodes and considers a predictor model $\{w_1, w_4\} \rightarrow \{w_3\}$.
% \item[DM+TO] This is the same estimator as the previous one, with the assumption that the missing node $w_2$ is measurable (oracle assumption). Therefore, it considers a predictor model $\{w_1, w_2, w_4\} \rightarrow \{w_3\}$ where $w_2$ is known. Note that in order to avoid biased target module estimates in the direct method framework, we need $w_2$ to be measured and included as one of the predictor inputs \cite{Dankers&etal_TAC:16}.
 \end{description}

We run $50$ independent Monte Carlo experiments where the data are generated using known reference signals $r_2(t)$ and $r_4(t)$ that are realizations of white noise with unit variance. The number of data samples is $N$ = 150. The noise sources $e_1(t)$, $e_2(t)$, $e_3(t)$ and $e_4(t)$ have variance 0.05, 0.08, 0.5, 0.1, respectively. We assume that we know the model order of $G_{31}^0(q)$.  For the method DM+TO+M, we solve a 2-input/1-output MISO identification problem with $w_1(t)$ and $w_4(t)$ as inputs, which should lead to a biased target module estimate \cite{Dankers&etal_TAC:16}. As for DM+TO, with the assumption that the missing node $w_2$ is measurable (oracle assumption), we solve a 3-input/1-output MISO identification problem with $w_1(t)$, $w_2(t)$ and $w_4(t)$ as inputs in order to compare the results of our developed method. For both these cases we consider that the model orders of all the modules in the MISO structure are known. Analogously, EBDM considers a 3-input/1-output MISO identification problem, while EBDM+M solves a 2-input/1-output MISO identification problem. For MC-EBDM, MC-EBDMA, EBDM, EBDM+M we choose $l = 15$.

\begin{figure}[h]
	\centering
	\hspace*{-0.5cm}
	\includegraphics[scale=0.23]{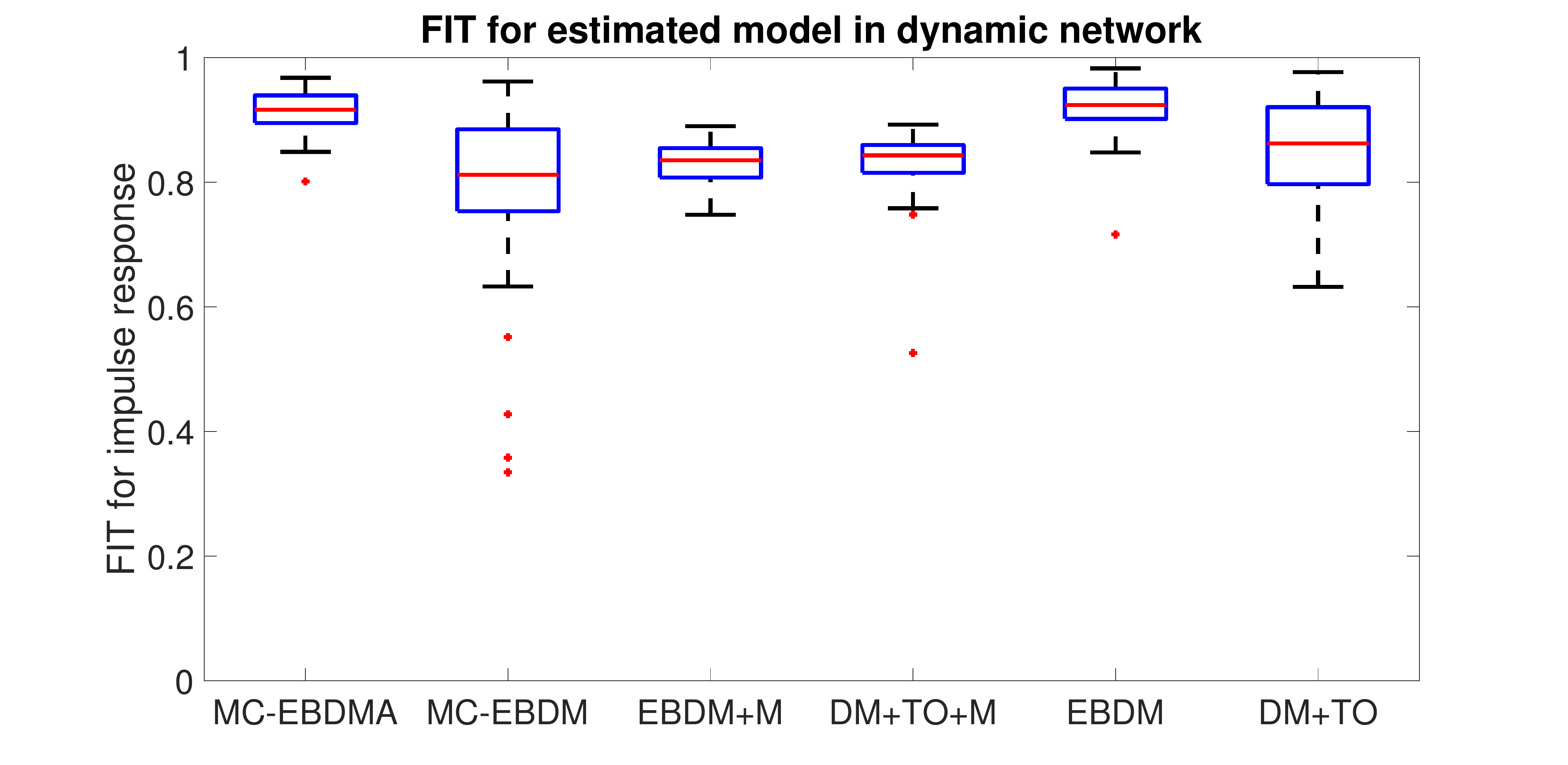}
	\caption{Box plot of the fit of the impulse response of $\hat{G}_{31}$ obtained using different methods. EBDM and DM+TO assumes that the missing node $w_2$ is measurable (oracle assumption) and use it for the estimation.}
	\label{fig:boxplotimp}
\end{figure}

\begin{figure}[h]
	\centering
	\hspace*{-0.5cm}
	\includegraphics[scale=0.23]{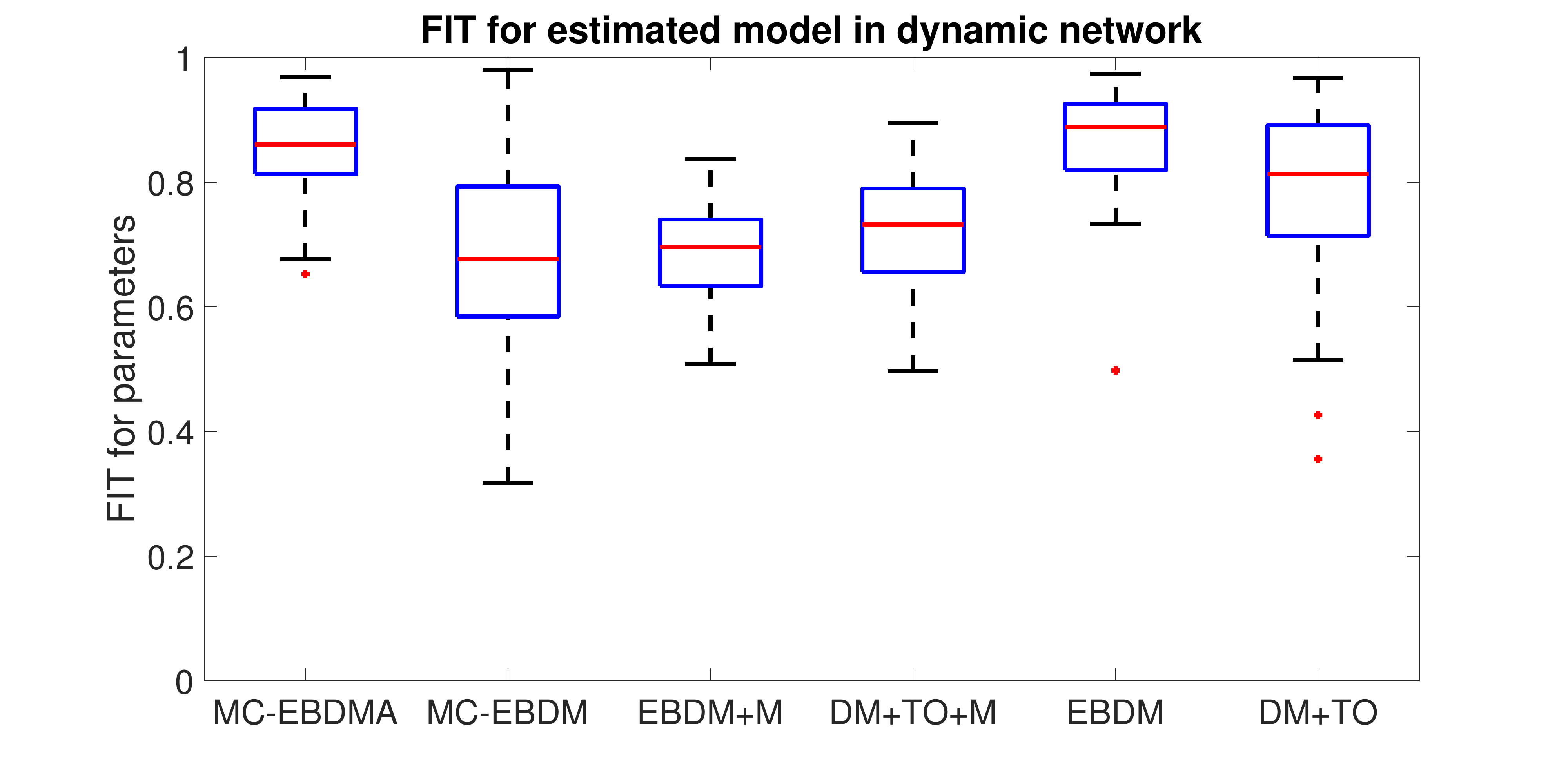}
	\caption{Box plot of the fit of the parameters of $\hat{G}_{31}$ obtained using different methods.}
	\label{fig:boxplotpara}
\end{figure}

To evaluate the performance of the methods, we use the standard goodness-of-fit metric,
\beq
\textrm{Fit}_{imp} = 1 - \frac{{\norm{g_{ji}^0 - \hat g_{ji}}}_2}{{\norm{g_{ji}^0 - \bar g_{ji}}}_2}, \nonumber \ ; \ \textrm{Fit}_{\theta} = 1 - \frac{{\norm{\theta^0 - \hat \theta}}_2}{{\norm{\theta^0 - \bar \theta}}_2}, \nonumber
\eeq
where $\textrm{Fit}_{imp}$ and $\textrm{Fit}_{\theta}$ are the fit of the estimated impulse response and estimated parameters of the target module respectively. $g_{ji}^0$ is the true value of the impulse response of $G_{ji}^0$, $\hat g_{ji}$ is the impulse response of the estimated target module, $\bar g_{ji}$ is the sample mean of $g_{ji}^0$, $\theta^0$ is the true parameter of the target module, $\hat \theta$ is the estimated value of the parameter and $\bar \theta$ is the sample mean of $\theta^0$. The box plots of the fit of the impulse response and box plots of the fit of the parameters of $G_{31}(q)$ are shown in Figure \ref{fig:boxplotimp} and \ref{fig:boxplotpara} respectively for the above mentioned methods.

\begin{figure}
	\centering
	\hspace*{-1cm}
	\includegraphics[scale=0.25]{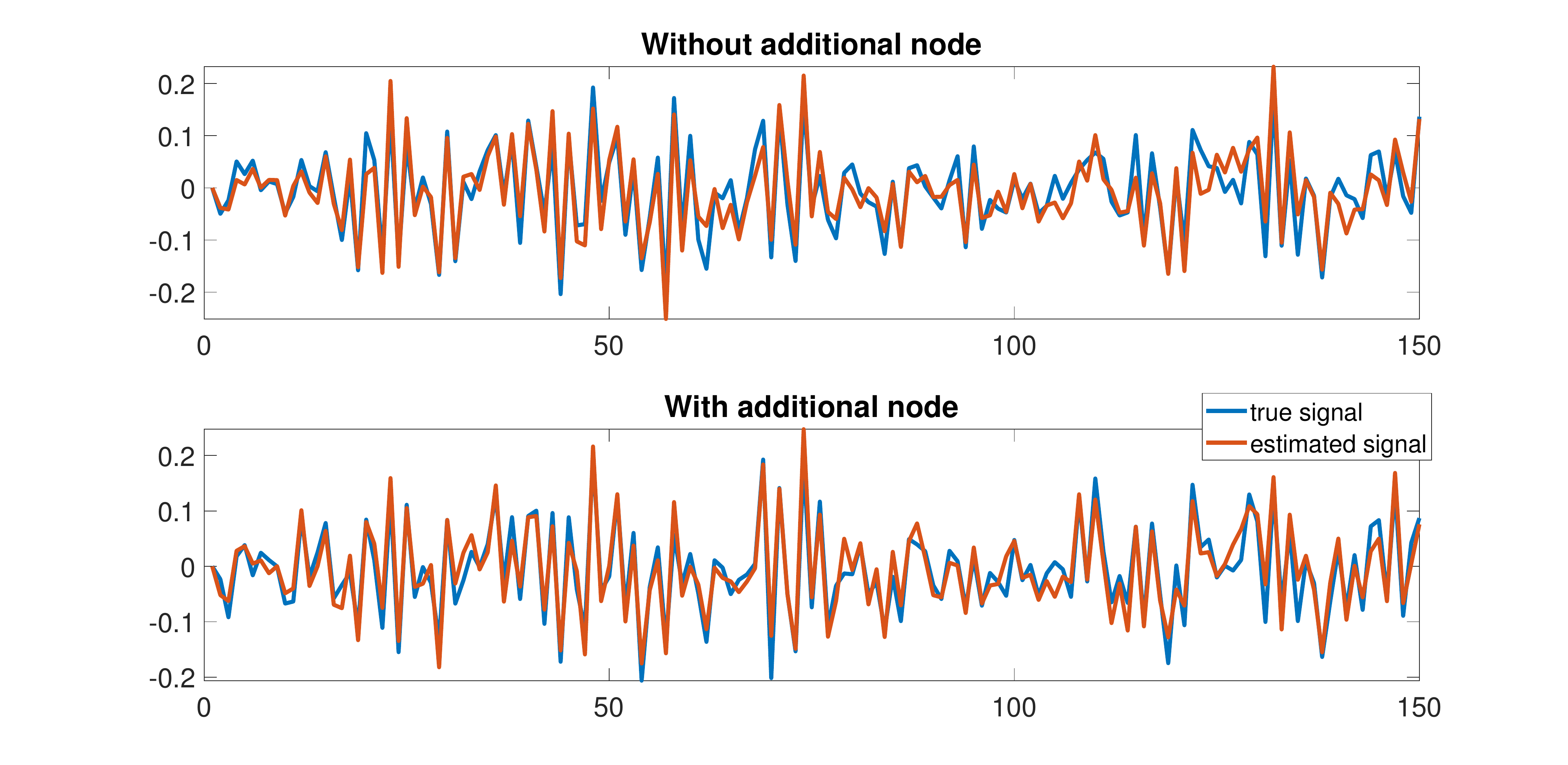}
	\caption{Re-constructed missing observation $w_2$ (normalized) signal for one MC simulation using MC-EBDM (top) and MC-EBDMA (bottom), compared with the measured value (blue) of $w_2$ over $N = 150$ data points.}
	\label{fig:reconstructed}
\end{figure}

\begin{figure}
	\centering
	\hspace*{-1cm}
	\includegraphics[scale=0.25]{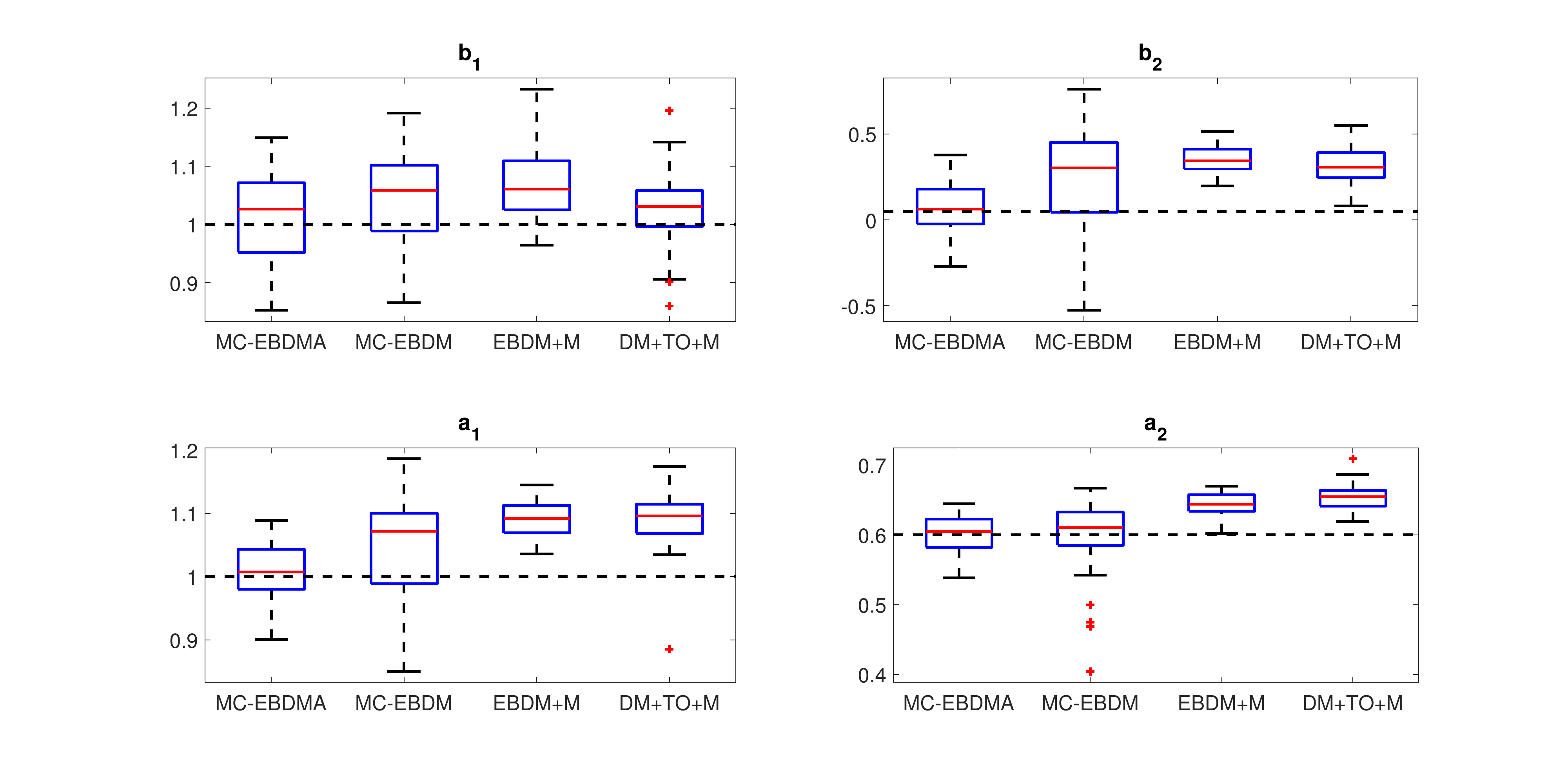}
	\caption{Box plot of the estimate for each parameter obtained from 50 MC simulations using different identification methods.}
	\label{fig:boxploteachpara}
\end{figure}

It can be noted that both MC-EBDMA, despite considering $w_2$ to be non-measured, achieve significantly better fit than the other methods that do not consider $w_2$ to be known (i.e. DM+TO+M and EBDM+M). Comparing with methods that consider $w_2$ to be known (i.e. DM+TO and EBDM), the novel estimator MC-EBDMA performs better than the direct method; furthermore, MC-EBDMA achieves a fit comparable to the fit obtained by the oracle EBDM. Also, the performance of MC-EBDM is poor compared to other methods, and thus shows the importance of including additional node(s) (i.e. MC-EBDMA). In Figure \ref{fig:reconstructed}, we show the re-constructed signal $w_2$ for one MC simulation using MC-EBDM and MC-EBDMA. It can be seen that considering additional nodes aid better reconstruction of the missing node observation and provides better estimates.
Figure \ref{fig:boxploteachpara} shows the box plot of each parameter estimates of $G_{31}$.
It is evident that the developed method MC-EBDMA provides smaller bias and greatly reduced variance under the case of missing node observations compared to the other methods. The reduction in variance is attributed to the regularization approach used in this method. Figure \ref{fig:boxploteachpara} again shows the importance of adding additional node(s) (i.e. MC-EBDMA) since the MC-EBDM has a larger variance compared to the other methods. Therefore, compared to the available methods for network identification, the developed framework stands out as an effective method that can handle the situation of missing node observations by reconstructing the node signal and offer reduced variance estimates. Considering the situation of large-sized networks, the developed method also circumvents the model order selection step that is required for the standard direct methods, which leads to computational burden and inaccurate results.

\section{Conclusions}\label{sec:conclusion}

Identification of a local module in a dynamic network typically leads to a MISO or MIMO identification problem where all node signals that are part of the predictor model are assumed to be available from measurements.In this paper, we have introduced an effective method for dealing with the situation that one or more of the required available node signals are not measured, but reconstructed as part of a Bayesian parameter estimation problem, by using a data-augmentation strategy. By using a regularized kernel-based approach, the method also circumvents an extensive  model order selection step for all the modules in the predictor model, and offers a substantial reduction in the number of parameters to be estimated. For the optimization, an EM algorithm is followed where the E-step is solved by the particle approximation MCEM approach, by drawing samples from the posterior using Gibbs sampler. The introduced framework is not dependent on this particular choice, but allows alternative steps as well, like the Variational Bayes EM (VBEM) \cite{Beal2003}.
Numerical simulations performed with a dynamic network example shows promising results, and illustrate the potentials of the developed method to reconstruct the missing node observations and provide reduced variance estimates.

\bibliographystyle{plain}        % Include this if you use bibtex
\bibliography{Paul_Dynamic_Networks_Library}
\appendix
%
% %%%%%%%%%%%%%%%%%%%%%%%%%%%%%%%%%%%%%%%%%%%%%%%%%%%%%%%%%%%%%%%%%%%
% %
\section{Proof of Proposition \ref{prop2}}
Let us first consider the conditional distribution $p(w_m|w_{\bar\smY},s, p, f)$. We first write,
\beqr\label{eq:wm}
w_m & = & \mathbf R_{\backslash m} b + W_mb_m + u_m + \xi_m \, , \\
w_{\bar\smY} & = &  \mathbf W_{\smD_{\backslash m}} g_{\backslash m} + \mathbf W_mg_m + \mathbf W_{ji}g_{ji} + u_{\bar\smY} + \xi_{\bar\smY} \, ,\label{eq:wj}
\eeqr
where $\mathbf W_{\smD_{\backslash m}}$ is constructed after excluding $W_m$ in the matrix $\mathbf W_{\smD}$ and $g_{\backslash m}$ is the vector constructed after excluding $s_m$ and $f_m$ in $g$. Here, $g_m = \begin{bmatrix}
s_m^\top & f_m^\top
\end{bmatrix}^\top$ and $\mathbf W_m = blkdiag(W_m,W_m)$.
Grouping terms in \eqref{eq:wm} and \eqref{eq:wj} we get,
\beqr\label{eq:wm_new}
w_m & = & \mu_3 + \bar B_mw_m + \xi_m \, , \\
w_{\bar\smY} & = &  \mu_1 + \mu_2w_m  + \xi_{\bar\smY} \,
,\label{eq:wj_new}
\eeqr
where $\mu_1 = \mathbf W_{\smD_{\backslash m}} g_{\backslash m} + \mathbf W_{ji}g_{ji} + u_{\bar\smY}$, $\mu_2 = \begin{bmatrix}
\bar S_m \\ \bar F_m
\end{bmatrix}$, $\mu_3 = \mathbf R_{\backslash m} p + u_m$ and $\Sigma_{\bar\smY} = blkdiag(\sigma_j^2I_N,\sigma_a^2I_N)$.
By the law of conditional expectation and ignoring terms independent of $w_m$, we write
\beqr
&&\log p(w_m|w_{\bar\smY},b, s,\! f) \cong \log p(w_{\bar\smY}, w_m| s, b, f) \nonumber\\
&&\quad\quad\quad\quad\cong - \frac{1}{2}||\begin{bmatrix}
	w_{\bar\smY} \\ w_m
\end{bmatrix} - \begin{bmatrix}
\mu_1 \\ \mu_3
\end{bmatrix} - \begin{bmatrix}
\mu_2 \\ \bar B_m
\end{bmatrix}w_m||^2_{\Sigma^{-1}}  \nonumber\\
&&\quad\quad\quad\quad\cong - \frac{1}{2}||w_m||^{2}_{P_w^{-1}} + w_m^\top P_w^{-1} \mu_w \, ,
\eeqr
where $\bar \mu_1 = \begin{bmatrix}
	\mu_1 \\ \mu_3
\end{bmatrix}$, $\bar \mu_2 = \begin{bmatrix}
\mu_2 \\ \bar B_m
\end{bmatrix}$, $\Sigma^{-1} = \begin{bmatrix}
\Sigma_{11} & \Sigma_{12}\\
\Sigma_{21} & \Sigma_{22}
\end{bmatrix}^{-1} = \begin{bmatrix}
\Lambda_{11} & \Lambda_{12}\\
\Lambda_{21} & \Lambda_{22}
\end{bmatrix}$, $P_w = \left({\bar\mu_2^\top\Sigma^{-1}\bar\mu_2} + \Lambda_{22} - 2\begin{bmatrix}
\Lambda_{21} & \Lambda_{22}
\end{bmatrix}\bar\mu_2\right)^{-1}$, $\mu_w = P_w(\bar\mu_2^\top \begin{bmatrix}
\Lambda_{11} & \Lambda_{12}
\end{bmatrix}^\top w_{\bar\smY} + \begin{bmatrix}
\Lambda_{21} & \Lambda_{22}
\end{bmatrix} \bar\mu_1 -\Lambda_{21}w_{\bar\smY} - \bar \mu_2^\top \Sigma^{-1}\bar\mu_1)$. The above log density is quadratic and represents a Gaussian distribution with covariance $P_w$ and mean $\mu_w$.

%%%%%%%%%%%%%% p(s|others)%%%%%%%%%%%%%%%%%%%%
Let us now consider the conditional distribution $p(s|w_{\bar\smY},w_m, b, f)$.
%We first write,
%\beqr
%w_{\smY}  = \mu_4 + \begin{bmatrix}
%\mathbf W \\ \mathbf 0_N
%\end{bmatrix}s + \xi_{\smY} \, ,
%\eeqr
%where $\mu_4  = \mathbf W_{ji}g_{ji} + \begin{bmatrix}
%\mathbf 0_N \\ \mathbf Q
%\end{bmatrix}f + r_{\smY}$ and
By the law of conditional expectation and ignoring terms independent of $s$, we write
\beqr
&&\log p(s|w_{\bar\smY},w_m,b, f) \cong \log p(w_{\bar\smY}, w_m| s, b, f) + p(s) \nonumber\\
&&\quad\quad\quad\quad\cong - \frac{1}{2\sigma_j^2}||w_{\smY} - \bar \mu_4 - \bar{\mathbf W}s||^2_{\Sigma^{-1}} \!\!-\!\! \frac{1}{2}||s||^2_{\mathbf K_{1}^{-1}} \, , \nonumber \\
&&\quad\quad\quad\quad\cong - \frac{1}{2}||s||^{2}_{P_s^{-1}} + s^\top P_s^{-1} \mu_s \, ,
\eeqr
where $\bar \mu_4  = \begin{bmatrix}
\mathbf W_{ji} \\ \mathbf 0_N
\end{bmatrix}g_{ji} +
\bar{\mathbf Q}f + \bar{\mathbf R}b + \begin{bmatrix}
u_{\bar\smY} \\ u_m
\end{bmatrix}$, $\bar{\mathbf W} = \begin{bmatrix}
\mathbf W \\ \mathbf 0_N \\ \mathbf 0_N
\end{bmatrix}$, $\bar{\mathbf Q} = \begin{bmatrix}
\mathbf 0_N \\ {\mathbf Q}\\\mathbf 0_N
\end{bmatrix}$, $\bar{\mathbf R} = \begin{bmatrix}
\mathbf 0_N \\\mathbf 0_N \\ {\mathbf R}
\end{bmatrix}$,
$P_s = \left(\mathbf K_1^{-1} + \bar{\mathbf W}^\top{\Sigma^{-1}}\bar{\mathbf W}\right)^{-1}$, $\mu_s = P_s\bar{\mathbf W}^\top{\Sigma^{-1}}\left(w_{\smY} - \bar \mu_4\right)$ and $\mathbf 0_N$ is a zero matrix with $N$ rows.
The above log density is quadratic and represents a Gaussian distribution with covariance $P_s$ and mean $\mu_s$.

%%%%%%%%%%%%%% p(f|others)%%%%%%%%%%%%%%%%%%%%
Let us now consider the conditional distribution $p(f|w_{\bar\smY},w_m, b, s)$.
%We first write,
%\beqr
%w_{\smY}  = \mu_6 + \begin{bmatrix}
%	\mathbf 0_N \\ \mathbf Q
%\end{bmatrix}f + \xi_{\smY} \, ,
%\eeqr
%where $\mu_6  = \mathbf W_{ji}g_{ji} + \begin{bmatrix}
%\mathbf W \\ \mathbf 0_N
%\end{bmatrix}s + r_{\smY}$.
By the law of conditional expectation and ignoring terms independent of $s$, we write
\beqr
&&\log p(f|w_{\bar\smY},w_m,b, s) \cong \log p(w_{\bar\smY}, w_m| s, b, f) + p(f) \nonumber\\
&&\quad\quad\quad\quad \cong - \frac{1}{2\sigma_j^2}||w_{\smY} - \bar \mu_6 - \bar{\mathbf Q}f||^2_{\Sigma^{-1}} \!\!-\!\! \frac{1}{2}||f||^2_{\mathbf K_{3}^{-1}} \, , \nonumber \\
&&\quad\quad\quad\quad \cong - \frac{1}{2}||f||^{2}_{P_f^{-1}} + f^\top P_f^{-1} \mu_f \, ,
\eeqr
where $\bar \mu_6  = \begin{bmatrix}
\mathbf W_{ji} \\ \mathbf 0_N
\end{bmatrix}g_{ji} +
\bar{\mathbf W}s +
\bar{\mathbf R}b + \begin{bmatrix}
u_{\bar\smY} \\ u_m
\end{bmatrix}$,
$P_f = (\mathbf K_3^{-1} + \bar{\mathbf Q}^\top{\Sigma^{-1}}\bar{\mathbf Q})^{-1}$, and
$\mu_f = P_f\bar{\mathbf Q}^\top{\Sigma^{-1}}\left(w_{\smY} - \bar \mu_6\right)$.
The above log density is quadratic and represents a Gaussian distribution with covariance $P_f$ and mean $\mu_f$.

%%%%%%%%%%%%%% p(b|others)%%%%%%%%%%%%%%%%%%%%
Let us now consider the conditional distribution $p(b|w_{\bar\smY},w_m, s, f)$. By the law of conditional expectation and ignoring terms independent of $p$, we write
\beqr
\log p(b|w_{\bar\smY}, w_m, s, f) &\cong& \log p(w_{\bar\smY}, w_m| s, b, f) + p(b) \nonumber\\
&\cong& - \frac{1}{2}||w_{\smY} - \bar\mu_5 - \bar{\mathbf R}b||^2_{\Sigma_{\smY}^{-1}} \!\!-\!\! \frac{1}{2}||b||^2_{\mathbf K_{2}^{-1}} \nonumber\\
&\cong& - \frac{1}{2}||p||^{2}_{P_p^{-1}} + p^\top P_p^{-1} \mu_p \, ,
\eeqr
where $\bar\mu_5  = \begin{bmatrix}
\mathbf W_{\smD} \\ \mathbf 0_N
\end{bmatrix}\begin{bmatrix}
s \\ f
\end{bmatrix} + \begin{bmatrix}
\mathbf W_{ji} \\ \mathbf 0_N
\end{bmatrix}g_{ji} + \begin{bmatrix}
u_{\bar\smY} \\ u_m
\end{bmatrix}$, $P_b = (\mathbf K_2^{-1} + \bar{\mathbf R}^\top{\Sigma^{-1}}\bar{\mathbf R})^{-1}$, and
$\mu_b = P_b\bar{\mathbf R}^\top{\Sigma^{-1}}\left(w_{\smY} - \bar \mu_5\right)$.
The above log density is quadratic and represents a Gaussian distribution with covariance $P_b$ and mean $\mu_b$.

\vspace{-0.1cm}
\section{Proof of Proposition \ref{prop3a}}
From \eqref{eq:optisplit} we have,
\beqr\label{eq:C}
C &=& \argmax \!\bigg[\sum_{k}{Q_s}_k^{(n)}({\lambda_k^s},{\beta_k^s}) \!+\! \frac{1}{M}  \sum_{i = 1}^{M}\log p(\bar s_{m}^{(i,n)}; \beta_{m}^s)\bigg] \nonumber
\eeqr
with $k = \{j, c_1, \dots, c_p, jk_1, \dots, jk_p\}$ where $c_1,\,\ldots,\,c_p$ and $k_1,\,\ldots,\,k_p$ are the elements of the set $\mathcal{D}_j^w \backslash \{i,m \}$ and $\mathcal{D}_j^r\backslash \{j\}$ respectively, and
\beqr
&&{Q_{s_k}}^{(n)}({\lambda_k^s},{\beta_k^s}) = \frac{1}{M}\sum_{i = 1}^{M} \log p(\bar s_{k}^{(i,n)}; \lambda_{k}^s,\beta_{k}^s) \nonumber\\
&& \cong \frac{1}{M}\sum_{i = 1}^{M}-\log[\det({\lambda_k^s}{K_{\beta_k^s}})] \!\!-\! \mathrm{tr}({({\lambda_k^s}{K_{\beta_k^s}})}\!^{-1}{\bar s_{k}^{(i,n)}}{\bar s_{k}^{\top(i,n)}}) \nonumber \\
&& \cong - \log[\det({\lambda_k^s}{K_{\beta_k^s}})] - \mathrm{tr}({({\lambda_k^s}{K_{\beta_k^s}})}^{-1}\hat{\mathbf S}_k^{(n)}) \nonumber\\
&& \quad\quad\quad\quad\quad\quad\quad\quad\quad\quad\quad - \hat s_k^{\top(n)}{(\lambda_k^s{K_{\beta_k^s}})}^{-1}\hat s_k^{(n)}.  \label{eq:102}
\eeqr
Next, the proof follows the procedure used in \cite{Bottegaletal_Autom:16}. We partially differentiate \eqref{eq:102} with respect to $\lambda_{k}^s$ and equate to zero to get the $\lambda_{k}^{s*}$ expression. Substituting this $\lambda_{k}^{s*}$ in \eqref{eq:102} we get the expression for \eqref{eq:101} using which we obtain ${\hat\beta}_{{k}}^{s(n+1)}$. Equation~\eqref{eq:13.1} is the expression of $\lambda_{k}^{s*}$ after substituting ${\hat\beta}_{{k}}^{s(n+1)}$.

\section{Proof of Proposition \ref{prop3b}}
Considering $\argmax \frac{1}{M}  \sum_{i = 1}^{M}\log p(\bar s_{m}^{(i,n)}; \beta_{m}^s)$ in \eqref{eq:C} and expanding it as in \eqref{eq:102} where $\lambda_{m}^s = 1$, we get the result of the proposition.

\section{Proof of Proposition \ref{prop5}}
From \eqref{eq:optisplit} we have,
\beqr
A
% &=& \argmin_{\theta,\bar\Sigma}\!\!\frac{1}{M}\sum_{i = 1}^{M}\bigg[\sum_{t = 1}^{N}\log\det(\bar\Sigma) \nonumber \\
% &&\quad\quad\quad\quad+ \sum_{t=1}^N \bar \varepsilon^{(i,n)}(t)^\top\bar\Sigma^{-1}\bar\varepsilon^{(i,n)}(t)\bigg] \nonumber\\
&=& \argmin_{\theta,\sigma_j^2}\!\!\frac{1}{M}\sum_{i = 1}^{M}\bigg[N\log\sigma_j^2 +  \frac{1}{\sigma_j^2}\bar \varepsilon_j^{(i,n)\top}\bar\varepsilon_j^{(i,n)})\bigg] \nonumber\\
&+&\argmin_{\tilde\Sigma}\!\!\frac{1}{M}\sum_{i = 1}^{M}\bigg[\sum_{t = 1}^{N}\log\det(\tilde\Sigma) \nonumber\\
&&\quad\quad\quad\quad+ \sum_{t=1}^N \mathrm{tr}(\tilde\Sigma^{-1}\bar\varepsilon^{(i,n)}(t)\bar \varepsilon^{(i,n)}(t)^\top)\bigg] \nonumber\\
&=& A_1 + A_2.
\nonumber
\eeqr
We now write,
\beqr
A_2 &=& \argmin_{\tilde\Sigma}\!\!\bigg[\sum_{t = 1}^{N}\log\det(\tilde\Sigma) \nonumber\\
&&\quad\quad+ \sum_{t=1}^N \mathrm{tr}\left(\tilde\Sigma^{-1}\left[\hat \varepsilon^{(n)}(t)\hat \varepsilon^{(n)}(t)^\top + \hat{\mathbf E}^{(n)}(t)\right]\right)\bigg]. \nonumber
\eeqr
For the optimization problem $A_2$, we can follow the similar reasoning as the maximum likelihood proof in \cite{Astrom_Autom:1980}, which yields the result of the proposition for estimating parameters in $\tilde \Sigma$.
% This is done by differentiating the above cost function with respect to the elements of $\bar\Sigma^{-1}$ and using the relations
% \begin{align}
%     \log \det\tilde\Sigma &= - \log\det\tilde\Sigma^{-1} \nonumber\\
%     \det P &= \sum_i p_{ij}p^{ij} \nonumber\\
%     (P^{-1})_{ij} &= p^{ij}/\det P
% \end{align}
% where $p^{ij}$ denotes the cofactor of the $ij^{th}$ element $p_{ij}$ of the matrix $P$.

Now considering $A_1$, we can write
\begin{align}\label{eq:14.1}
 A_1 & \!=\! \argmin_{\theta,\sigma_j^2}\!\!\bigg[N\log\sigma_j^2 \!+\! \frac{1}{\sigma_j^2}\frac{1}{M}\sum_{i = 1}^{M} \bar \varepsilon_j^{(i,n)\top}\bar\varepsilon_j^{(i,n)}\bigg].
\end{align}
We notice that the optimum with respect to $\theta$ does not depend on the optimal value of $\sigma_j^2$. Then, we can first update $\theta$ and then use its optimal value to update $\sigma_j^2$. In order to find ${\hat\theta}^{(n)}$, $\sigma_j^2$ is fixed to $\hat\sigma_j^{2(n)}$ and substituted in Eq.~\eqref{eq:14.1}. After substitution, the terms that are independent of $\theta$ can be removed from the objective function since it becomes a constant. Then we get,
\begin{align}
\hat \theta^{(n+1)} & = \argmin_{\theta}\sum_{i = 1}^{M} \bar \varepsilon_j^{(i,n)\top}\bar\varepsilon_j^{(i,n)}
\end{align}
Let us define $\breve{s}^{(i,n)} \in \mathbb{R}^{N}$ be a vector such that, if $N \leq l$, $\breve{s}^{(i,n)}$ is the vector of first $N$ elements of $\bar s^{(i,n)}$ and if $N > l$, $\breve{s}^{(n)}$ is a vector with the first $l$ elements equal to $\bar s^{(i,n)}$ and the remaining ones equal to 0. Let $\breve{S}^{(i,n)}$, $\breve{W}_i \in \mathbb{R}^{N\times N}$ be the Toeplitz matrix of $\hat s^{(n)}$ and $\bar{{\mathbf w}}_i$ respectively.
Then
$$\bar{\mathcal{X}}^{(i,n)} = {\begin{bmatrix}
W_j - R_j & {W_{c_1}} & \dots & {R_{k_p}}
\end{bmatrix}}, \hspace{10pt} \bar{\mathcal{Y}}^{(i,n)} = \breve{S}^{(i,n)}\breve{W}_i.$$
We now re-write,
$\bar{\mathbf W}^{(i,n)}\bar s^{(i,n)} = \bar{\mathcal{X}}^{(i,n)}\bar s^{(i,n)} +  G_{\theta}\bar W_{i}\bar s_j^{(i,n)}$ $= \bar{\mathcal{X}}^{(i,n)}\bar s^{(i,n)} + \bar{\mathcal{Y}}^{(i,n)}g_{ji}$. Therefore,
\begin{align}
\hat \theta^{(n+1)} & = \argmin_{\theta}\sum_{i = 1}^{M} \bigg[-2w_{j}^\top W_{ji}g_{ji} -2w_{j}^\top\bar{\mathcal{Y}}^{(i,n)}g_{ji} \nonumber\\
&+ g_{ji}^\top W_{ji}^\top W_{ji}g_{ji} \!+\! 2g_{ji}^\top W_{ji}^\top r_j \!+\! 2g_{ji}^\top W_{ji}^\top\bar{\mathcal{X}}^{(i,n)}\bar s^{(i,n)} \nonumber\\
&+ 2 g_{ji}^\top W_{ji}^\top\bar{\mathcal{Y}}^{(i,n)}g_{ji} + 2r_j^\top\bar{\mathcal{Y}}^{(i,n)}g_{ji} \nonumber\\
&+ 2\bar s^{(i,n)\top} \bar{\mathcal{X}}^{(i,n)\top}\bar{\mathcal{Y}}^{(i,n)}g_{ji} +g_{ji}^\top\bar{\mathcal{Y}}^{(i,n)\top} \bar{\mathcal{Y}}^{(i,n)}g_{ji}
\bigg]. \nonumber
\end{align}
Defining
$\displaystyle
\hat A^{(n)} = \sum_{i = 1}^{M}[W_{ji}^\top W_{ji} + 2W_{ji}^\top \bar{\mathcal{Y}}^{(i,n)}+ \bar{\mathcal{Y}}^{(i,n)\top} \bar{\mathcal{Y}}^{(i,n)}],
$ and
\begin{equation*}
\begin{split}
\hat{b}^{(n)} &= \sum_{i = 1}^{M}\Big[w_{j}^\top W_{ji} + w_{j}^\top\bar{\mathcal{Y}}^{(i,n)} - r_j^\top W_{ji}- r_j^\top\bar{\mathcal \mY}^{(i,n)}\\
&- \bar{s}^{(i,n)\top}\bar{\mathcal X}^{(i,n)\top} W_{ji} - \bar s^{(i,n)\top} \bar{\mathcal{X}}^{(i,n)\top}\bar{\mathcal{Y}}^{(i,n)}\Big]^\top,
\end{split}
\end{equation*}
we get the statement of the proposition for updating $\hat\theta^{(n+1)}$.

In order to find $\hat\sigma_j^{2(n)}$, $\theta$ is fixed to ${\hat\theta}^{(n+1)}$ and substituted in Eq.~\eqref{eq:14.1}. After substitution, $A_1(\sigma_j^2,\hat{\theta}^{(n+1)})$ is differentiated w.r.t. $\sigma_j^2$ and equated to zero which leads to the result of the proposition.

\end{document}